\documentclass[twocolumn,showpacs,superscriptaddress,preprintnumbers,amsmath,amssymb,fleqn]{revtex4}

\usepackage{graphicx}
\usepackage{dcolumn}
\usepackage{bm}
\usepackage{color}
\usepackage{ulem}

\newcommand{\blue}[1]{\textcolor{blue}{#1}}

\newcommand{\bq}{\begin{equation}}
\newcommand{\eq}{\end{equation}}
\newcommand{\bqa}{\begin{eqnarray}}
\newcommand{\eqa}{\end{eqnarray}}
\newcommand{\nn}{\nonumber \\}

\def\be     {\begin{equation}}
\def\ee     {\end{equation}}
\def\bwt     {\begin{widetext}}
\def\ewt    {\end{widetext}}
\def\bea        {\begin{eqnarray}}
\def\eea        {\end{eqnarray}}
\def\bnn    {\begin{eqnarray*}}
\def\enn    {\end{eqnarray*}}

\begin{document}

\title{Dissipationless mechanism of skyrmion Hall current in
double-exchange ferromagnets}

\author{Ki-Seok Kim}

\affiliation{Asia Pacific Center for Theoretical Physics, Hogil
Kim Memorial building 5th floor, POSTECH, Pohang, Gyeongbuk
790-784, Korea} \affiliation{Department of Physics, POSTECH,
Pohang, Gyeongbuk 790-784, Korea}

\author{Shigeki Onoda}
\affiliation{Condensed Matter Theory Laboratory, RIKEN, 2-1,
Hirosawa, Wako 351-0198, Japan}
\date{\today}

\begin{abstract}
We revisit a theory of skyrmion transport in ferromagnets. On a
basis of an effective U(1) gauge theory for spin-chirality
fluctuations in double-exchange ferromagnets, we derive an
expression for the velocity of a skyrmion core driven by the dc
electric field. We find that mutual feedback effects between
conduction electrons and localized spins give rise to Chern-Simons
terms, suggesting a dissipationless mechanism for the skyrmion
Hall current. A conventional description of the current-induced
skyrmion motion, appearing through the spin transfer torque and
scattering events, is reproduced in a certain limit of our
description, where the Chern-Simons terms are not fully
incorporated. Our theory is applicable to not only metallic but
also insulating systems, where the purely topological and
dissipationless skyrmion Hall current can be induced in the
presence of an energy gap.
\end{abstract}

\maketitle

\section{Introduction}

Soliton dynamics in the presence of fermions has been one of the
most fundamental issues in various fields of physics. It plays a
central role in domain-wall dynamics in conducting polymers
\cite{Conducting_Polymer} and vortex dynamics in superconductors
in the field of condensed-matter physics
\cite{Vortex_Dynamics_Review}. It is also relevant to a
confinement and baryon dynamics in high-energy physics
\cite{Soliton_High_Energy}. Skyrmions \cite{Skyrmion_Original}
were shown to appear as elementary excitations in the quantum Hall
system \cite{Skyrmion_IQHE}, and the interplay between skyrmions
and spin-wave excitations was also studied
\cite{Skyrmion_Magnon_IQHE}. Actually, skyrmion excitations have
been observed in the cold atom system
\cite{Skyrmion_Condensation}, and their crystallization has also
been observed in both two and three dimensional helical magnets
\cite{MnSi_Skyrmion_Crystal,Film_Skyrmion_Crystal}.
%
%

Advances in spintronics have promoted intensive and extensive
studies on domain-wall dynamics in magnetic systems for its
application to the magnetic memory device
\cite{Current_Driven_Motion1,DomainWall_Dynamics_Review}. In
particular, spin-polarized electric currents allow for an
efficient control of domain-wall dynamics because of the spin
torque transfer. A conventional theoretical approach to this
phenomenon is based on the Landau-Lifshitz-Gilbert equation, that
is, the equation of motion for a single spin. In this formalism
the so-called Gilbert damping term is introduced in a
phenomenological manner \cite{DomainWall_Dynamics_Review} to take
account of the dissipation in the spin dynamics, caused by the
coupling to the conduction electrons and the relativistic
spin-orbit coupling. It reveals that the spin current in the
itinerant ferromagnet drives a domain-wall motion to the
longitudinal direction \cite{DomainWall_Dynamics_Review}. The spin
dynamics in nanoscale magnets has also been studied by introducing
collective coordinates \cite{Collective_Coordinate}, such as
positions of domain wall and magentic vortices. Then, it has been
argued that the spin current generates half magnetic vortices
and/or anti-vortices which can exhibit a nontrivial motion,
including the transverse motion in the presence of the Gilbert
damping. In these previous approaches spin currents are given,
thus feedback effects of the spin dynamics on the electron
dynamics have not been considered.

In this paper, we develop a theory of skyrmion transport in
double-exchange ferromagnets, where conduction electrons interact
with localized spins via the Hund's-rule coupling. An essential
aspect of our study lies in a mutual feedback effect between
electron spin currents and skyrmion currents, resulting in the
dissipationless skyrmion Hall current of the topological origin,
which should be distinguished from the dissipative skyrmion Hall
current replying on the scattering events
\cite{DomainWall_Dynamics_Review}.
%
Actually, a coupling between the spin current and a skyrmion motion has
also been discussed in Refs.
\cite{Berry_Phase_Electricity,Berry_Phase_Electricity_Feedback,Wong:09,Wong:10},
where topological magnetic textures generate electric currents via
the Berry-phase induced electro-motive force, which produces
feedback on the magnetization dynamics via a spin-transfer torque.
In these pioneering works, however, they considered the
dissipative mechanism of the skyrmion Hall current but not the
dissipationless mechanism.

In Sec.~\ref{sec:gauge theory}, resorting to the CP$^{1}$
representation for the localized spin, we derive an effective
field theory for itinerant electrons and bosonic spinons, which
interact via gauge fluctuations representing the spin chirality.
Based on this effective field theory for the strong Hund's-rule
coupling limit \cite{DomainWall_Dynamics_Review}, we derive the
Maxwell-Chern-Simons equations for both internal U(1) gauge
fluctuations and external electromagnetic fields. The emergence of
Chern-Simons terms is ascribed to mutual feedbacks between
itinerant electrons and skyrmions. Focusing on the center-of-mass
motion of the skyrmion, we obtain the velocity of the skyrmion
core in terms of the dc electric field. The Chern-Simons terms
induce the dissipationless skyrmion Hall current normal to the
applied electric field. For comparison with previous theories, we
show in Sec.~\ref{sec:compare} that our U(1) gauge-theory
formulation for skyrmion dynamics reproduces the dissipative
skyrmion Hall current in the Landau-Lifshitz-Gilbert-equation
approach.

In fact, the topological contribution of the dissipationless
skyrmion Hall current scales with the skyrmion density, and thus
vanishes in the thermodynamic limit when a single skyrmion is
considered. However, we find that the Rashba spin-orbit coupling
for conduction electrons produces a finite skyrmion Hall current
in the thermodynamic limit, allowing for an observation of
this intrinsic skyrmion Hall effect. This could be realized at the
surface of three-dimensional topological insulators \cite{Qi_TI}
when a ferromagnet is deposited. Recently, a dissipationless
mechanism for magnetization switching was proposed in the
topological surface state, where the Chern-Simons term plays an
essential role \cite{Franz_TI}.

\section{U(1) gauge theory for single skyrmion dynamics}
\label{sec:gauge theory}

\subsection{An effective Maxwell-Chern-Simons Lagrangian}

We start from an effective U(1) gauge-field formulation of a
two-dimensional double exchange model with the Rashba spin-orbit
coupling, where itinerant electrons interact with localized spins
via the ferromagnetic Hund's-rule coupling $J_H$. This is
described with the following partition function $Z$ and the
Lagrangian density ${\cal L}$;
\begin{widetext} \bqa
&& Z = \int D \psi_{\sigma} D z_{\sigma} D a_{\mu}
\delta(|z_{\sigma}|^{2} - 1) \delta(\partial_{r}a_{r}) e^{ -
\int_{0}^{\beta} d \tau \int d^{2} r {\cal L}_{eff} } , ~~~~~
{\cal L} = {\cal L}_{B} + {\cal L}_{\psi} + {\cal L}_{z} , \nn &&
{\cal L}_{B} = 2 i S a_{\tau} , ~~~~~ {\cal L}_{\psi} =
\psi_{\sigma}^{\dagger} (
\partial_{\tau} - \mu_{r} - J_{H} S \sigma - i \sigma a_{\tau} - i A_{\tau})
\psi_{\sigma} + t |[\partial_{\mathbf{r}} - i \sigma
a_{\mathbf{r}} - i A_{\mathbf{r}} - i \sigma (\lambda_{so}/t)
z_{\alpha}^{\dagger} \sigma^{\mathbf{r}}_{\alpha\beta} z_{\beta}]
\psi_{\sigma}|^{2} , \nn && {\cal L}_{z} = \rho_{s}
z_{\sigma}^{\dagger}(\partial_{\tau} - i a_{\tau})z_{\sigma} + t
\rho_{s} |[\partial_{\mathbf{r}} - i a_{\mathbf{r}} - i
(\lambda_{so}/t) z_{\alpha}^{\dagger}
\sigma^{\mathbf{r}}_{\alpha\beta} z_{\beta}]z_{\sigma}|^{2} .
\label{EFT} \eqa
\end{widetext}
This effective field theory can be derived along the concept of
Ref. \cite{Lee_Wen_Nagaosa}. The derivation is given in Ref.
\cite{Kim_Kim} as well as in Appendix A. The underlying mechanism
to justify this formulation is that the spin dynamics of itinerant
electrons instantaneously follows that of localized spins in the
strong Hund's-rule coupling limit as far as the dynamics of
localized spins are much slower than that of itinerant electrons
\cite{DomainWall_Dynamics_Review}. Below we explain the physical
meaning of each term.

${\cal L}_{B}$ represents the single-spin Berry phase, resulting
from the curved nature of the SU(2) spin manifold. This term is
indispensable for reproducing the skyrmion dynamics that has been
obtained in the Landau-Lifshitz-Gilbert equation approach
\cite{DomainWall_Dynamics_Review}.

${\cal L}_{\psi}$ describes the dynamics of itinerant electrons,
where $\psi_{\sigma}$ and $z_{\alpha}$ represent the fermionic
field for itinerant electrons and the CP$^{1}$ spinon field for
localized spins, respectively. As mentioned above, spins of
itinerant electrons follow those of localized electrons in the
strong Hund's-rule coupling limit. This constraint provides
dynamics of itinerant electrons with an effective internal flux,
originating from the curvature in the spin space. Namely, their
orbital motion is affected by an effective Aharonov-Bohm phase or
the Berry-phase connection, described by an internal U(1) gauge
field $a_{\mu}$ with $\mu = \tau, x, y$. Physically, this gauge
field represents spin chirality fluctuations, and couples to
electrons with opposite signs of coupling constants for ``spin" up
and down. In contrast, $A_\mu$ represents the external
electromagnetic potential which linearly couples to electric
charge/current density of itinerant electrons. $\mu_{r}$ is the
chemical potential, which is determined to fix the total number of
itinerant electrons.
$t$ is the hopping energy of itinerant electrons. The last term
represents a spin vector potential, which originates from the
Rashba spin-orbit coupling $\lambda_{so}$. This provides an
interaction between the spin and the ``spin current", and thus
quenches the spin direction to that of the momentum or spin
current.

${\cal L}_{z}$ describes the dynamics of the CP$^{1}$ spinon field
$z_{\alpha}$ for localized spins, in other words, their
directional (angular) fluctuations.
In ferromagnets, this spinon dynamics produces the magnon
excitations, which exhibits the $\omega \propto k^{2}$ dispersion
relation in the SU(2) symmetric case. We have introduced the spin
density of itinerant electrons, $\rho_{s} = \Bigl\langle
\sum_{\sigma} \sigma \psi_{\sigma}^{\dagger} \psi_{\sigma}
\Bigr\rangle$. We have an additional term generated by the Rashba
spin-orbit coupling $\lambda_{so}$.

Note that the skyrmion configuration creates a non-trivial
background potential for the Berry gauge connection $a_{\mu}$ and
thus a fictitious internal magnetic field in the $z$ direction.
This affects the dynamics/transport of itinerant electrons.
Indeed, even if the skyrmion is static, it produces the anomalous
Hall current of itinerant electrons \cite{ye,AHE}. Actually, this
is one side of the mutual feedback effect between the skyrmion and
the fermionic matter. On the other side, the topologically induced
anomalous Hall current is accompanied by the dissipationless
skyrmion Hall current, when the skyrmions are depinned intrinsic
objects, as we will show later.

The skyrmion motion can be uncovered from an effective action for
both U(1) Berry gauge fields $a_{\mu}$ and electromagnetic fields
$A_{\mu}$, integrating over electrons $\psi_{\sigma}$ and spinons
$z_{\sigma}$ in the skyrmion background. We separate the Berry
gauge field into two pieces which correspond to its classical
configuration and quantum-fluctuation part, respectively. The
classical configuration of the Berry gauge field is determined
from an equation of motion for spinons, where dynamics of spinons
is taken into account classically. The skyrmion solution of the
spinon field gives rise to an effective magnetic field for
electrons, given by the following relation of $a_{\mu}^{c} = -
\frac{i}{2} [z_{\sigma}^{c \dagger} (\partial_{\mu}
z_{\sigma}^{c}) - (\partial_{\mu} z_{\sigma}^{c \dagger})
z_{\sigma}^{c}]$, where the superscript $c$ denotes ``classical".
Inserting the effective magnetic field into the Schrodinger
equation for electrons, we construct the space of wave functions,
well fitted to the skyrmion potential. Then, we can integrate over
electrons, and expand the resulting logarithmic term up to the
second order for gauge fluctuations. The whole procedure is shown
in section IV.
%
%
This gives rise to not only the Maxwell Lagrangian ${\cal L}_{M}$
but also the spatially dependent Chern-Simons action ${\cal
L}_{CS}$ for the gauge-field dynamics,
\begin{widetext} \bqa
&& {\cal L}_{eff} = {\cal L}_B+{\cal L}_M+{\cal L}_{CS} , \nn &&
{\cal L}_{M} = \frac{1}{2} \left(
\begin{array}{cc} \delta a_{i} & A_{i} \end{array} \right)  \left(
\begin{array}{cc} \sigma_{ss} |\partial_{\tau}| + \chi_{ss} ( - \partial^{2})
& \sigma_{sc} |\partial_{\tau}| + \chi_{sc} ( -
\partial^{2}) \\ \sigma_{sc} |\partial_{\tau}| +
\chi_{sc} ( - \partial^{2}) & \sigma_{cc} |\partial_{\tau}| +
\chi_{cc} ( - \partial^{2})
\end{array} \right) P_{ij}^{T} \left(
\begin{array}{c} \delta a_{j} \\ A_{j} \end{array} \right) , \nn && {\cal L}_{CS} = i
\frac{\Theta_{ss}(\mathbf{x}-\mathbf{X})}{2\pi}
\epsilon_{\mu\nu\lambda} \delta a_{\mu}
\partial_{\nu} \delta a_{\lambda} + i \frac{\Theta_{sc}(\mathbf{x}-\mathbf{X})}{\pi}
\epsilon_{\mu\nu\lambda} \delta a_{\mu}
\partial_{\nu} A_{\lambda}
+ i \frac{\Theta_{cc}(\mathbf{x}-\mathbf{X})}{2\pi}
\epsilon_{\mu\nu\lambda} A_{\mu}
\partial_{\nu} A_{\lambda} , \label{EFT_CS}
\eqa
\end{widetext}
where $\delta a_{\mu}$ represents the quantum-fluctuation part.

In the Maxwell Lagrangian ${\cal L}_{M}$, $\sigma_{ss}$,
$\sigma_{cc}$, $\sigma_{sc}$, and $\chi_{ss}$, $\chi_{cc}$,
$\chi_{sc}$ are conductivities and diamagnetic susceptibilities,
associated with spin-current--spin-current,
charge-current--charge-current, and spin-current--charge-current
correlation functions, respectively. Electrons are assumed to be
in the diffusive regime, resulting in the $z = 2$ dynamics for
gauge fluctuations, where $z$ is the dynamical exponent to
represent the dispersion relation, $\omega \propto k^{z}$.
$P_{ij}^{T}$ is the projection operator for transverseness of the
gauge dynamics, given by $P_{ij}^{T} = \delta_{ij} +
\partial_{i} \partial_{j}/(-\partial^{2})$, where $i, j = x, y$
and $- \partial^{2} = - \partial_{x}^{2} - \partial_{y}^{2}$.
Dynamics of the temporal part $\delta a_{\tau}$ can be neglected
in the low energy limit because such fluctuations are gapped and
decoupled with spatial fluctuations in the Coulomb gauge.

In the Chern-Simons action ${\cal L}_{CS}$,
$\Theta_{ss}(\mathbf{x}-\mathbf{X})$ denotes a local spinon-Hall
conductance, given by the transverse spin-current--spin-current
correlation function. Namely, it describes the spin Hall current
generated by the magnetic field gradient.
$\Theta_{cc}(\mathbf{x}-\mathbf{X})$ is a local charge-Hall
conductance, given by the transverse
charge-current--charge-current correlation function.
$\Theta_{sc}(\mathbf{x}-\mathbf{X})$ is a local spin-Hall
conductance, given by the transverse spin-current--charge-current
correlation function. $\mathbf{X}$ represents the spatial
coordinate of the skyrmion core, which should be distinguished
from that of the fields, i.e., $\mathbf{x}$. Actually, the
skyrmion contribution to the charge Hall current decays with the
distance $|\mathbf{x}-\mathbf{X}|$ from the skyrmion core. In
particular, the charge Hall current vanishes at the long distance
if the relativistic spin-orbit coupling $\lambda_{so}$ is absent.
All coefficients in this effective field theory will be found in
section IV.

Based on Eq. (\ref{EFT_CS}), we investigate the skyrmion dynamics
under an external dc electric field applied along the $x$
direction. Since we are interested only in the transport
properties in the linear response to the applied dc electric
field, it is sufficient to treat the constant velocity of the
skyrmion core. To facilitate the calculation on the coupling
between itinerant electrons and localized spins, it is convenient
to introduce the frame moving with the skyrmion core at the
origin.
In this skyrmion moving frame, the time derivative and
the time component of the gauge field are transformed as
\bqa &&
\partial_{\tau} \longrightarrow \partial_{\tau} - \mathbf{v}_{r}
\cdot {\boldsymbol{\partial}_{r}} , ~~~~~ \delta a_{\tau}
\longrightarrow \delta a_{\tau} - \mathbf{v}_{r} \cdot \delta
\mathbf{a}_{r} , \label{Transformation} \eqa where
$\mathbf{v}_{r}$ is the constant skyrmion velocity driven by the
external electric field, which will be determined
self-consistently below. Note that we implicitly ignore the
modification of the shape of the static single-skyrmion
configuration, which requires more careful self-consistent
treatment of the coupling between itinerant electrons and
localized spins, but does not spoil the topological origin of our
skyrmion Hall current totally.


\subsection{Skyrmion dynamics under electric field}




Taking the derivative of ${\cal L}_{eff}$ with respect to $A_\mu$
and $\delta a_\mu$, we obtain an equation of motion for the U(1)
Berry gauge field and that for the electromagnetic field,
respectively,
\bqa \chi_{s} \partial^{2} \delta a_{i} + \chi_{cs}
\partial^{2} A_{i}
&=& 2 (S - M) v_{i} \nn &+& \sigma_{ss} \delta e_{i} -
\frac{\Theta_{ss} }{ \pi} \epsilon_{ij} \delta e_{j} -
\frac{\Theta_{sc} }{2\pi} \epsilon_{ij} E_{j} , \nn \chi_{c}
\partial^{2} A_{i} + \chi_{cs} \partial^{2} \delta a_{i}
&=& - \rho_{el} v_{i} \nn &+& \sigma_{cc} E_{i} -
\frac{\Theta_{cc} }{\pi} \epsilon_{ij} E_{j} - \frac{\Theta_{sc}
}{2\pi} \epsilon_{ij} \delta e_{j} , \label{Maxwell_CS} \eqa
where the imaginary time has been replaced with the real time.
%
%
The terms linearly proportional to the skyrmion velocity $v_{i}$
originate from the Berry-phase term in the moving frame. $M =
\Bigl\langle \sum_{\sigma} \sigma \psi_{\sigma}^{\dagger}
\psi_{\sigma} + \rho_{s} \sum_{\sigma} z_{\sigma}^{\dagger}
z_{\sigma} \Bigr\rangle$ corresponds to the magnetization density,
which effectively reduces the coefficient of the Berry phase term
and suppresses that of the skyrmion velocity. $\rho_{el} =
\Bigl\langle \sum_{\sigma} \psi_{\sigma}^{\dagger} \psi_{\sigma}
\Bigr\rangle$ is the charge density. $E_{j}$ is an external
electric field and $\delta e_{j} = \epsilon_{j \mu \nu}
\partial_{\mu} \delta a_{\nu}$ is an internal electric field.

Physics of these Maxwell-Chern-Simons equations can be understood
as follows. Recalling the structure of the Maxwell equation, one
can construct two constituent equations, which relate ``spin'' and
``charge'' currents with both internal and external electric
fields, \bqa \sum_{\sigma} \sigma J_{i\sigma}^{\psi}  &=&
\sigma_{ss} \delta e_{i} - \frac{\Theta_{ss} }{ \pi} \epsilon_{ij}
\delta e_{j} - \frac{\Theta_{sc} }{2\pi} \epsilon_{ij} E_{j} , \nn
\sum_{\sigma} J_{i\sigma}^{\psi} &=& \sigma_{cc} E_{i} -
\frac{\Theta_{cc} }{\pi} \epsilon_{ij} E_{j} - \frac{\Theta_{sc}
}{2\pi} \epsilon_{ij} \delta e_{j} , \label{Constituent_Eq} \eqa
where $J_{i\sigma}^{\psi}$ represents the current of itinerant
electrons with the spin index $\sigma$ flowing in the
$j$-direction, given by \bqa J_{j\sigma}^{\psi} &=& - i t
[\psi_{\sigma}^{\dagger} \{ (\partial_{j} - i \sigma
(\lambda_{so}/t)
z_{\alpha}^{\dagger}\sigma_{\alpha\beta}^{j}z_{\beta})
\psi_{\sigma}\} \nn &-& \{ (\partial_{j} + i \sigma
(\lambda_{so}/t)
z_{\alpha}^{\dagger}\sigma_{\alpha\beta}^{j}z_{\beta})
\psi_{\sigma}^{\dagger}\} \psi_{\sigma}]
%
. \label{Current_SO} \eqa
%
%
%
The spin conductivity $\sigma_{ss}$ vanishes in the paramagnetic
phase without the spin-orbit coupling, while in the ferromagnetic
phase with the Zeeman splitting $J_{H}S $, it is finite and the
spin current of itinerant electrons is generated by the external
electric field $E_{i}$. Equation (\ref{Constituent_Eq})
generalizes the standard constituent relation in metals, where
dynamics of conduction electrons are in the diffusive regime,
introducing the Chern-Simons contribution into the equation, which
plays an essential role for the mutual feedback effect between
skyrmions and itinerant electrons. The presence of the
Chern-Simons term in the constituent equation confirms the
anomalous Hall effect of itinerant electrons in the metallic
ferromagnet, as discussed before.


%
%
%


It is straightforward to solve coupled equations
(\ref{Maxwell_CS}) when the spatial dependence for both gauge
fields is neglected. Performing integration of $\int d x \int d y$
in Eq. (\ref{Maxwell_CS}), we obtain
\bqa && - 2(S-M) v_{i} = \sigma_{ss} \delta e_{i} -
\frac{\sigma_{ss}^{H}}{\pi L^{2}} \epsilon_{ij} \delta e_{j} -
\frac{\sigma_{sc}^{H}}{2\pi L^{2}} \epsilon_{ij} E_{j} , \nn &&
\rho_{el} v_{i} = \sigma_{cc} E_{i} - \frac{\sigma_{cc}^{H}}{\pi
L^{2}} \epsilon_{ij} E_{j} - \frac{\sigma_{sc}^{H}}{2\pi L^{2}}
\epsilon_{ij} \delta e_{j} , \label{Skyrmion_Motion} \eqa
where \bqa && \sigma_{ss}^{H} = \int dx \int dy
\Theta_{ss}(\mathbf{x}) , \nn && \sigma_{cc}^{H} = \int dx \int dy
\Theta_{cc}(\mathbf{x}) , \nn && \sigma_{sc}^{H} = \int dx \int dy
\Theta_{sc}(\mathbf{x}) \label{Hall_Conductances} \eqa are Hall
conductivities with $L$ being the linear spatial dimension of the
system.

Coupled equations (\ref{Skyrmion_Motion}) describe both the
internal electric field $\delta e_{i}$ and the skyrmion velocity
$v_{i}$ as a function of the external electric field $E_{i}$. We
find the following expression for the internal electric field \bqa
&& \left(
\begin{array}{c} e_{x} \\ e_{y} \end{array} \right) = \Bigl( \frac{\sigma_{sc}^{H}}{2\pi L^{2}} \Bigr)^{-2} \Bigl\{ \rho_{el}
\left( \begin{array}{cc} 0 & \frac{\sigma^{H}_{sc}}{2\pi L^{2}} \\
- \frac{\sigma^{H}_{sc}}{2\pi L^{2}} & 0 \end{array} \right)
\left( \begin{array}{c} v_{x} \\ v_{y}
\end{array} \right) \nn && +
\left( \begin{array}{cc} - \frac{\sigma^{H}_{sc}}{2\pi L^{2}} \frac{\sigma^{H}_{cc}}{\pi L^{2}} & - \frac{\sigma^{H}_{sc}}{2\pi L^{2}} \sigma_{cc} \\
\frac{\sigma^{H}_{sc}}{2\pi L^{2}} \sigma_{cc} & -
\frac{\sigma^{H}_{sc}}{2\pi L^{2}} \frac{\sigma^{H}_{cc}}{\pi
L^{2}} \end{array} \right) \left(
\begin{array}{c} E_{x} \\ 0 \end{array}
\right) \Bigr\} . \label{epsilon} \eqa Inserting this expression
into Eq. (\ref{Skyrmion_Motion}), we obtain the skyrmion velocity
as a function of the external electric field $E_{x}$,
\begin{widetext} \bqa && v_{x} = \frac{ \Bigl( (S-M)
\sigma^{H}_{sc} + \rho_{el} \sigma^{H}_{ss} \Bigr) \Bigl(
\sigma_{ss} \sigma^{H}_{cc} + \sigma_{cc} \sigma^{H}_{ss} \Bigr) +
\rho_{el} \sigma_{ss} \Bigl( \pi^{2} L^{4} \sigma_{ss} \sigma_{cc}
- \sigma^{H}_{ss} \sigma^{H}_{cc} + \frac{\sigma^{H
2}_{sc}}{4}\Bigr)}{\Bigl( (S-M) \sigma^{H}_{sc} + \rho_{el}
\sigma^{H}_{ss} \Bigr)^{2} + \pi^{2} L^{4} (\rho_{el}
\sigma_{ss})^{2} } E_{x} , \nn && v_{y} = \frac{ - \frac{1}{\pi
L^{2}} \Bigl( (S-M) \sigma^{H}_{sc} + \rho_{el} \sigma^{H}_{ss}
\Bigr)\Bigl( \pi^{2} L^{4} \sigma_{ss} \sigma_{cc} -
\sigma^{H}_{ss} \sigma^{H}_{cc} + \frac{\sigma^{H 2}_{sc}}{4}
\Bigr) + \rho_{el} \sigma_{ss} \Bigl( \pi^{2} L^{4} \sigma_{ss}
\sigma_{cc} - \sigma^{H}_{ss} \sigma^{H}_{cc} \Bigr)}{\Bigl( (S-M)
\sigma^{H}_{sc} + \rho_{el} \sigma^{H}_{ss} \Bigr)^{2} + \pi^{2}
L^{4} (\rho_{el} \sigma_{ss})^{2} } E_{x} .
\label{Skyrmion_Velocity} \eqa
\end{widetext}

To understand the above expression, we consider two limiting
cases. First, we take the limit of $\sigma_{ss}^{H} =
\sigma_{cc}^{H} = \sigma_{sc}^{H} = 0$, resulting in \bqa && v_{x}
= \frac{ \sigma_{cc} }{ \rho_{el} } E_{x} , ~~~~~ v_{y} = \frac{
\sigma_{cc} }{ \rho_{el} } E_{x} . \label{SkV_Full_Dissipation}
\eqa
%
%
The skyrmion current is driven to not only the same direction as
the applied electric field but also the orthogonal direction
corresponding to the Hall motion.
%
%

Second, we take another limit of $\sigma_{ss} = \sigma_{cc} =
\sigma_{sc} = 0$, corresponding to an insulator. Then, we find
\bqa && v_{x} = 0 , \nn && v_{y} = \frac{1}{\pi L^{2}} \frac{
\sigma^{H}_{ss} \sigma^{H}_{cc} - \frac{\sigma^{H 2}_{sc}}{4} }{
(S-M) \sigma^{H}_{sc} + \rho_{el} \sigma^{H}_{ss} } E_{x} .
\label{SkV_Insulator} \eqa
%
%
This is a remarkable result. Although the skyrmion Hall current
induced by the electric field vanishes in the thermodynamic limit,
the nature of the skyrmion Hall current is dissipationless. We
cannot find any coefficients associated with conductivity, giving
rise to dissipation. This certainly originates from the
Chern-Simons terms, an essential feature of the
``self-consistent'' treatment.

It is interesting to observe that the dissipationless skyrmion
Hall current may not vanish in the thermodynamic limit if the
spin-orbit interaction is introduced. The Hall coefficient without
the spin-orbit coupling is proportional to the density of
skyrmions, thus $\sim L^{-2}$ in the case of the single skyrmion.
As a result, Hall conductivities of $\sigma_{ss}^{H}$,
$\sigma_{cc}^{H}$, and $\sigma_{sc}^{H}$ in Eq.
(\ref{SkV_Insulator}) are constants even after the spatial
integration. However, the spin-orbit interaction can give rise to
a finite value for the Chern-Simons coefficients in the
thermodynamic limit, thus its integral value corresponding to
$\sigma_{ss}^{H}$, $\sigma_{cc}^{H}$, and $\sigma_{sc}^{H}$ will
be proportional to $L^{2}$. Then, we obtain the dissipationless
skyrmion Hall current in ferromagnetic insulators with the
spin-orbit coupling.

\subsection{Discussion : Dissipationless skyrmion Hall current
in the surface state of three dimensional topological insulators}

We discuss the dissipationless skyrmion Hall current in the
surface state of three dimensional topological insulators, where
magnetic impurities are deposited.
%
%
An effective field theory for surface Dirac electrons with
localized spins is given by \bqa {\cal S} = {\cal S}_{B} + \int
d^{3} \boldsymbol{x} \Bigl\{ \bar{\psi} \Bigl( i
\boldsymbol{\hat{D}} - m \boldsymbol{\vec n} \cdot
\boldsymbol{\vec \tau} \Bigr) \psi + \frac{1}{2e^{2}}
(\epsilon_{\mu\nu\lambda} \partial_{\nu} A_{\lambda})^{2} \Bigr\}
. \nn \label{EFT_Dirac} \eqa ${\cal S}_{B}$ is the single-spin
Berry phase term to appear from the coherent-state representation
for the localized spin $\boldsymbol{\vec n}$ in the path-integral
quantization. $\psi$ represents the surface Dirac fermion in the
irreducible representation, and the covariant derivative is
$i\boldsymbol{\hat{D}} \equiv \gamma_{\mu} (i \partial_{\mu} +
A_{\mu})$, where $\gamma_{\mu}$ with $\mu = \tau, x, y$ is two by
two Dirac matrices and $A_{\mu}$ is an external electromagnetic
vector potential. $m > 0$ is an effective coupling constant
between surface Dirac fermions and deposited magnetic impurities.
Compared with the double exchange model, the only difference is
that non-relativistic electrons are replaced with Dirac fermions.

Integrating over Dirac electrons and performing the gradient
expansion for the resulting logarithmic term \cite{NLSM_Hopf}, one
finds the following effective action \bqa {\cal S}_{eff} &=& {\cal
S}_{B} + \int d^{3} \boldsymbol{x} \Bigl( \frac{m}{8\pi}
(\partial_{\mu} \boldsymbol{\vec n})^{2} + i A_{\mu} J_{\mu} \nn
&+& \frac{i}{4\pi} \epsilon_{\mu\nu\lambda} A_{\mu}\partial_{\nu}
A_{\lambda} + \frac{1}{2e^{2}} (\epsilon_{\mu\nu\lambda}
\partial_{\nu} A_{\lambda})^{2} \Bigr) + i \pi
\vartheta[\boldsymbol{\vec n}] , \nn \label{EFT_Dirac_CS} \eqa
where dynamics of localized spins is governed by the non-linear
$\sigma$ model with additional terms. The conserved current \bqa
&& J_{\mu} = \frac{1}{8\pi} \epsilon_{\mu\nu\lambda}
\boldsymbol{\vec n} \cdot
\partial_{\nu} \boldsymbol{\vec n} \times
\partial_{\lambda} \boldsymbol{\vec n} \nonumber \eqa
corresponds to the skyrmion current, discussed before. The minimal
coupling between the skyrmion current and the electromagnetic
field implies that a skyrmion carries an electric charge, where a
normalizable fermion zero mode exists on the topological soliton,
inducing a fermionic charge. On the other hand, the last term
shows the geometric phase, identified with the Hopf term, which
determines the statistics of the topological soliton and its spin
quantum number. These two terms are well known in the field
theory, referred as the quantum anomaly, where the induced
fermionic charge for a soliton is a phenomenon due to the local
anomaly while the topological phase is due to the global anomaly
\cite{Soliton_TextBook,NLSM_Hopf}. Mathematically speaking, the
existence of such topological terms is guaranteed by the
fundamental property of the elliptic operator, the Dirac operator
in the present case, called the Atiya-Singer index theorem
\cite{Index_Theorem}.

One can express the above effective field theory as follows,
resorting to the CP$^{1}$ representation, \bqa {\cal S}_{eff} &=&
{\cal S}_{B} + \int d^{3} \boldsymbol{x} \Bigl( \frac{1}{2g^{2}}
|(\partial_{\mu} - i a_{\mu})z_{\sigma}|^{2} + \frac{i}{2\pi}
\epsilon_{\mu\nu\lambda} A_{\mu} \partial_{\nu} a_{\lambda} \nn
&+& \frac{i}{4\pi} \epsilon_{\mu\nu\lambda} a_{\mu}
\partial_{\nu} a_{\lambda} + \frac{i}{4\pi}
\epsilon_{\mu\nu\lambda} A_{\mu}
\partial_{\nu} A_{\lambda} + \frac{1}{2e^{2}}
(\epsilon_{\mu\nu\lambda}
\partial_{\nu} A_{\lambda})^{2}\Bigr) \nn \label{EFT_Dirac_CS_CP1} \eqa with $g^{2}
\propto 1/m$, where the skyrmion current is $J_{\mu} =
\frac{1}{2\pi} \epsilon_{\mu\nu\lambda} \partial_{\nu}
a_{\lambda}$ and the topological phase is
$\vartheta[\boldsymbol{\vec n}] = \frac{1}{4\pi^{2}}
\epsilon_{\mu\nu\lambda} a_{\mu}
\partial_{\nu} a_{\lambda}$. This effective field theory for
dynamics of localized spins on the topological surface is
essentially the same as our previous effective field theory for
dynamics of localized spins in the presence of non-relativistic
electrons except for the fact that coefficients in topological
terms, i.e., Chern-Simons terms have finite values in the
thermodynamic limit. As a result, we find the dissipationless
skyrmion Hall current in the surface state of three dimensional
topological insulators, although surface Dirac electrons become
gapped and insulating due to time reversal symmetry breaking.
Furthermore, the electric charge of the skyrmion will give rise to
an additional contribution for the Hall voltage beyond that from
gapped Dirac fermions.

It seems to be clear that the dissipationless skyrmion Hall
current will survive on the topological surface state in the
thermodynamic limit. However, it is difficult to guarantee such a
phenomenon in the case of non-relativistic electrons with the
spin-orbit interaction. First of all, it should be noted that our
effective field theory [Eq. (\ref{EFT})] is valid only in the
limit of $J_{H} \gg t > \lambda_{so}$. In the $J_{H} \rightarrow
0$ limit $\rho_{s} = \Bigl\langle \sum_{\sigma} \sigma
\psi_{\sigma}^{\dagger} \psi_{\sigma} \Bigr\rangle \propto J_{H}
S$ will vanish. As a result, we lose terms to describe dynamics of
local spins in the effective Lagrangian. This results from the
U(1) projection [Eq. (A6)] of the original SU(2) effective theory
[Eq. (A5)]. On the other hand, the $J_{H} \rightarrow 0$ limit
recovers the Rashba model via an appropriate gauge transformation
in Eq. (A5). In the strong Hund-coupling limit the spin-orbit
interaction gives rise to an additional internal magnetic field
for itinerant electrons, as shown in Eq. (\ref{EFT}). Inserting
the skyrmion configuration into the spin-orbit induced gauge
field, we can see that the internal effective magnetic flux given
by the spin-orbit interaction, $\frac{1}{L^{2}} \int d^{2}
\boldsymbol{r} [\partial_{x} (z_{\alpha}^{\dagger}
\sigma^{y}_{\alpha\beta} z_{\beta}) -
\partial_{y} (z_{\alpha}^{\dagger} \sigma^{x}_{\alpha\beta}
z_{\beta})]$, decreases as $1/L^{2}$, where $L$ is the system
size. The effective magnetic flux via the gauge field of the
skyrmion configuration, $\frac{1}{L^{2}} \int d^{2} \boldsymbol{r}
(\partial_{x} a_{y} - \partial_{y} a_{x})$, is also proportional
to $1/L^{2}$ as discussed before. As a result, the spin-orbit
interaction does not allow the dissipationless skyrmion Hall
current to survive in the thermodynamic limit when $J_{H}$ is
assumed to be large in Eq. (\ref{EFT}).

In the $J_{H} \rightarrow 0$ limit the situation is more tricky.
As well known, the Rashba model ($J_{H} = 0$ in Eq. (A1)) shows
the spin Hall effect \cite{Sinova}, where disorder effects are not
taken into account. When ferromagnetic interactions are turned on,
a helical ordered state is expected to appear. The question is
whether the spin Hall effect survives or not in the helical
ordered state. When the spin Hall effect exists, the
dissipationless skyrmion Hall current will be observed in the
thermodynamic limit. Unfortunately, we do not have any definite
answer because the helical order will change the electron
dispersion of the Rashba model, which can spoil the spin Hall
effect. In particular, the ordering wave vector may be
incommensurate generically, making the problem much complicated.
We leave this interesting problem as a future work.

\section{Comparison with the Landau-Lifshitz-Gilbert equation approach}
\label{sec:compare}

The well known Landau-Lifshitz-Gilbert equation \bqa &&
\frac{\partial \boldsymbol{\vec S}}{\partial t} = \gamma
\vec{\mathbf{ B }} \times \boldsymbol{\vec S} - \frac{\alpha}{S}
\boldsymbol{\vec S} \times \frac{\partial \boldsymbol{\vec
S}}{\partial t} \label{LLG1} \eqa is generalized as \bqa
\frac{\partial \boldsymbol{\vec S}}{\partial t} &=& \gamma
\vec{\mathbf{ B }} \times \boldsymbol{\vec S} - \frac{\alpha}{S}
\boldsymbol{\vec S} \times \frac{\partial \boldsymbol{\vec
S}}{\partial t} \nn &-& \frac{a^{3}}{2e S} (\boldsymbol{\vec
j}_{s}\cdot\boldsymbol{\vec \nabla}) \boldsymbol{\vec S} -
\frac{a^{3}\beta}{2e S^{2}} [\boldsymbol{\vec S} \times
(\boldsymbol{\vec j}_{s}\cdot\boldsymbol{\vec \nabla})
\boldsymbol{\vec S} ] \label{LLG2} \eqa in the presence of spin
current \cite{DomainWall_Dynamics_Review}. The first term in Eq.
(\ref{LLG1}) is the standard precession term with $\gamma = g
\mu_{B} / \hbar > 0$, where $g = 2$ is the $g$-factor and
$\mu_{B}$ is the Bohr magneton. $\vec{\mathbf{ B }}$ is an
effective magnetic field. The second term in Eq. (\ref{LLG1}) is
the Gilbert damping term, phenomenologically introduced, where
$\alpha$ is a damping coefficient and $S$ is spin. The third term
in Eq. (\ref{LLG2}) describes the spin-transfer torque from the
spin current, where $\boldsymbol{\vec j}_{s}$ is the spin current
and $a$ is the Bohr radius. The fourth term in Eq. (\ref{LLG2})
represents another torque contribution, perpendicular to the
spin-transfer torque, argued to result from spin relaxation of
conduction electrons. This term is called the $\beta$ term due to
the coefficient $\beta$.

One can derive an equation of motion for domain walls, vortices,
and skyrmions from this generalized Landau-Lifshitz-Gilbert
equation, resorting to the collective coordinate method
\cite{Collective_Coordinate}. Suppose that the right hand side of
Eq. (\ref{LLG1}) or Eq. (\ref{LLG2}) can be derived from an
effective Hamiltonian $H_{eff}$. Then, one can write down an
equation of motion for spin dynamics as follows \bqa &&
\frac{\partial \boldsymbol{\vec S}}{\partial t} = -
\boldsymbol{\vec S} \times \frac{\delta H_{eff}}{\delta
\boldsymbol{\vec S}} . \nonumber \eqa Multiplying
$\boldsymbol{\vec S} \times \frac{\partial \boldsymbol{\vec
S}}{\partial \xi_{i}}$ to both sides, we obtain \bqa &&
\boldsymbol{\vec S} \cdot \frac{\partial \boldsymbol{\vec
S}}{\partial \xi_{i}} \times \frac{\partial \boldsymbol{\vec
S}}{\partial t} = - \frac{\partial \boldsymbol{\vec S}}{\partial
\xi_{i}} \cdot \frac{\delta H_{eff}}{\delta \boldsymbol{\vec S}} ,
\nonumber \eqa where the orthogonality condition $\boldsymbol{\vec
S} \cdot \frac{\partial \boldsymbol{\vec S}}{\partial \xi_{i}} =
0$ is used. $\xi_{i}$ is called the collective coordinate,
representing the core position of the skyrmion (domain wall or
vortex), where $i = x, y$. Integrating both sides over the two
dimensional space area $\mathbf{A}$, we reach the following
expression for the skyrmion dynamics \bqa && 4 \pi Q \sum_{j = x,
y} \epsilon_{ij} \frac{\partial \xi_{j}}{\partial t} = -
\int_{\mathbf{A}} d^{2} \mathbf{r} \frac{\delta H_{eff}}{\delta
\xi_{i}} , \eqa where $Q$ is an integer identified with the
skyrmion number \bqa && 4 \pi Q \epsilon_{ij} = \int_{\mathbf{A}}
d^{2} \mathbf{r} \boldsymbol{\vec S} \cdot \frac{\partial
\boldsymbol{\vec S}}{\partial \xi_{i}} \times \frac{\partial
\boldsymbol{\vec S}}{\partial \xi_{j}} . \eqa

The skyrmion or vortex dynamics has been studied in this
framework, starting from essentially the same effective action as
the present model Hamiltonian [Eq. (A1)] \bqa && S_{eff} = \int
d^{3} x \Bigl\{ c_{\sigma}^{\dagger} \Bigl( - i \partial_{t} -
\frac{\partial_{\mathbf{r}}^{2}}{2m} \Bigr) c_{\sigma} - J_{H}
\boldsymbol{\vec n} \cdot c_{\alpha}^{\dagger} \boldsymbol{\vec
\sigma}_{\alpha\beta} c_{\beta} \nn && + S \dot{\phi} (1 -
\cos\theta) + \frac{J S^{2}}{2} (\partial_{\mathbf{r}}
\boldsymbol{\vec n})^{2} + V(\boldsymbol{\vec n}) \Bigr\} ,
\nonumber \eqa where $\boldsymbol{\vec n} = (\sin \theta \cos
\phi, \sin \theta \sin \phi, \cos \theta)$ represents a spin
direction and $V(\boldsymbol{\vec n})$ introduces spin anisotropy.
$J$ is the ferromagnetic exchange coupling constant and
$\dot{\phi} = \partial_{t} \phi$.

Performing the same unitary transformation as Eq. (A3), one
obtains the same effective action as Eq. (A5) except for the
Rashba spin-orbit coupling term \bqa && S_{eff} = \int d^{3} x
\Bigl\{ \psi_{\sigma}^{\dagger} \Bigl( - i (\partial_{t} - i
a_{t}) - \frac{(\partial_{\mathbf{r}} - i a_{\mathbf{r}})^{2}}{2m}
\Bigr) \psi_{\sigma} \nn && - J_{H} \sigma \psi_{\sigma}^{\dagger}
\psi_{\sigma} + S \dot{\phi} (1 - \cos\theta) + \frac{J S^{2}}{2}
(\partial_{\mathbf{r}} \boldsymbol{\vec n})^{2} +
V(\boldsymbol{\vec n}) \Bigr\} , \nonumber \eqa where the Berry
gauge field is given by $a_{\mathbf{r}} = \boldsymbol{\vec
a}_{\mathbf{r}} \cdot \boldsymbol{\vec \sigma} = - i
\boldsymbol{U^{\dagger}}
\partial_{\mathbf{r}} \boldsymbol{U}$. As intensively discussed
before, an essential effect is given by the spin chirality
fluctuation (Berry gauge field) to the spin-current minimal
coupling term \bqa && H_{ST} = \int d^{3} x \boldsymbol{\vec
j}_{s} \cdot \frac{\boldsymbol{\vec \nabla} \phi}{2} (1 - \cos
\theta) , \nonumber \eqa where the gauge field is represented with
two angles.

Based on this effective action, one can derive an equation of
motion for the vortex or skyrmion dynamics \cite{Shibata} \bqa &&
\boldsymbol{\vec G} \times \bigl( \boldsymbol{\vec v}_{s} -
\partial_{t} \boldsymbol{\vec \xi} \bigr) = - \frac{\partial
U(\boldsymbol{\vec \xi})}{\partial \boldsymbol{\vec \xi}} - \alpha
\partial_{t} \boldsymbol{\vec \xi} . \label{LLG} \eqa $\boldsymbol{\vec G} =
\boldsymbol{\hat e}_{z} S \int d^{3} x \vec{\mathbf{n}} \cdot
(\partial_{x} \vec{\mathbf{n}} \times \partial_{y}
\vec{\mathbf{n}}) = 4 \pi S Q \boldsymbol{\hat e}_{z} $ expresses
the internal magnetic flux given by the skyrmion charge $Q$.
$\boldsymbol{\vec v}_{s}$ represents the velocity associated with
the spin current of  itinerant electrons. $U(\boldsymbol{\vec
\xi})$ can be regarded as a pinning potential, and the first term
in the right hand side describes the associated pinning force.
$\alpha$ is the Gilbert damping constant.

Comparing this equation of motion with Eq. (\ref{LLG2}), the
extended Landau-Lifshitz-Gilbert equation, we see that $\gamma
\vec{\mathbf{ B }} \times \boldsymbol{\vec S}$ and $-
\frac{a^{3}\beta}{2e S^{2}} [\boldsymbol{\vec S} \times
(\boldsymbol{\vec j}_{s}\cdot\boldsymbol{\vec \nabla})
\boldsymbol{\vec S} ]$ in Eq. (\ref{LLG2}) are not introduced in
this equation of motion. $\boldsymbol{\vec G} \times \partial_{t}
\boldsymbol{\vec \xi}$ is associated with $\frac{\partial
\boldsymbol{\vec S}}{\partial t}$, and $\boldsymbol{\vec G} \times
\boldsymbol{\vec v}_{s}$ corresponds to $\frac{a^{3}}{2e S}
(\boldsymbol{\vec j}_{s}\cdot\boldsymbol{\vec \nabla})
\boldsymbol{\vec S}$. $\alpha
\partial_{t} \boldsymbol{\vec \xi}$ results from
$\frac{\alpha}{S} \boldsymbol{\vec S} \times \frac{\partial
\boldsymbol{\vec S}}{\partial t}$.

Each term in our Maxwell-Chern-Simons approach [Eq.
(\ref{Maxwell_CS})] has its partner in Eq. (\ref{LLG}) except for
Chern-Simons terms proportional to $\Theta_{cc}$ and
$\Theta_{sc}$. Considering the following constituent equations
\bqa \sum_{\sigma} \sigma J_{i\sigma}^{\psi}  &=& \sigma_{ss}
\delta e_{i} - \frac{\Theta_{ss} }{ \pi} \epsilon_{ij} \delta
e_{j} , \nn \sum_{\sigma} J_{i\sigma}^{\psi} &=& \sigma_{cc} E_{i}
, \nonumber \eqa we obtain \bqa && v_{x} = \frac{ \rho_{el}
\sigma^{H 2}_{ss} \sigma_{cc} + \pi^{2} L^{4} \rho_{el}
\sigma_{ss}^{2} \sigma_{cc} }{ \rho_{el}^{2} \sigma^{H 2}_{ss} +
\pi^{2} L^{4} \rho_{el}^{2} \sigma_{ss}^{2} } E_{x} , \nn && v_{y}
= \frac{ - \pi L^{2} \rho_{el} \sigma^{H}_{ss} \sigma_{ss}
\sigma_{cc} + \pi^{2} L^{4} \rho_{el} \sigma_{ss}^{2} \sigma_{cc}
}{ \rho_{el}^{2} \sigma^{H 2}_{ss} + \pi^{2} L^{4} \rho_{el}^{2}
\sigma_{ss}^{2} } E_{x} \nonumber \eqa from our
Maxwell-Chern-Simons approach. On the other hand, Eq. (\ref{LLG})
gives rise to the following skyrmion motion \bqa && v_{x} =
\frac{(4\pi S Q)^{2} (\sigma_{cc}/\rho_{el})}{(4 \pi S Q)^{2} +
\alpha^{2}} E_{x} , \nn && v_{y} = - \frac{(4\pi S Q) \alpha
(\sigma_{cc}/\rho_{el})}{(4 \pi S Q)^{2} + \alpha^{2}} E_{x} ,
\nonumber \eqa where the drift velocity of the spin current is
given by $\boldsymbol{\vec v}_{s} = (\sigma_{cc}/\rho_{el}) E_{x}
\mathbf{\hat{x}}$. We note that the charge current in terms of the
$\psi_{\sigma}$ field contains the contribution of the spin
current in terms of the electron field $c_{\sigma}$. In this
respect the charge conductivity $\sigma_{cc}$ differs from the
actual electrical conductivity. Identifying $4 \pi S Q$ with
$\rho_{el} \sigma_{ss}^{H} / L^{2}$ and $\alpha$ with $\pi
\rho_{el} \sigma_{ss}$, the above expression becomes \bqa && v_{x}
= \frac{ \rho_{el}^{2} \sigma_{ss}^{H 2} \sigma_{cc} }{
\rho_{el}^{2} \sigma_{ss}^{H 2}
 + \pi^{2} L^{4} \rho_{el}^{2} \sigma_{ss}^{2}} E_{x} , \nn && v_{y}
= - \frac{\pi L^{2} \rho_{el} \sigma_{ss}^{H} \sigma_{ss}
\sigma_{cc} }{\rho_{el}^{2} \sigma_{ss}^{H 2} + \pi^{2} L^{4}
\rho_{el}^{2} \sigma_{ss}^{2}} E_{x} . \nonumber \eqa This result
coincides with the first two contributions in the
Maxwell-Chern-Simons-equation approach.

\section{Evaluation of Chern-Simons terms}

\subsection{Current-current correlation functions}

An important task is to show that the Chern-Simons coefficients
are non-vanishing. In this section we evaluate all current-current
correlation functions explicitly. Integrating over itinerant
electrons in ${\cal L}_{\psi}$ of Eq. (\ref{EFT}), we find the
local Chern-Simons action (${\cal L}_{CS}$) in Eq. (\ref{EFT_CS}).
Generally, we consider an effective action for gauge fluctuations
\begin{widetext} \bqa && \mathcal{S}_{eff}^{gauge} =
\int d \mathbf{x} \int d \mathbf{x}' \frac{1}{2} \left(
\begin{array}{cc} a_{\mu}(\mathbf{x}) & A_{\mu}(\mathbf{x})
\end{array} \right) \left(
\begin{array}{cc} \Pi_{\mu\nu}^{ss}(\mathbf{x},\mathbf{x}') & \Pi_{\mu\nu}^{sc}(\mathbf{x},\mathbf{x}')
\\ \Pi_{\mu\nu}^{cs}(\mathbf{x},\mathbf{x}') & \Pi_{\mu\nu}^{cc}(\mathbf{x},\mathbf{x}')
\end{array} \right) \left(
\begin{array}{c} a_{\nu}(\mathbf{x}') \\ A_{\nu}(\mathbf{x}')
\end{array} \right) , \eqa where the gauge kernel matrix is given
by \bqa \Pi_{\mu\nu}^{ss}(\mathbf{x},\mathbf{x'}) &=& t^{2}
\Bigl\langle T_{\tau} \Bigl\{ [\sigma J_{\sigma \mu}(\mathbf{x})]
[\sigma' J_{\sigma' \nu}(\mathbf{x}')] \Bigr\} \Bigr\rangle_{c} -
4 i t \lambda_{so} \Bigl\langle T_{\tau} \Bigl\{ [
n_{\mu}^{c}(\mathbf{x}) \rho_{\sigma}(\mathbf{x})] [\sigma'
J_{\sigma' \nu}(\mathbf{x}')] \Bigr\} \Bigr\rangle_{c} - \Bigl(
\mathbf{x} \leftrightarrow \mathbf{x}', \mu \leftrightarrow \nu
\Bigr) \nn &-& 16 \lambda_{so}^{2} \Bigl\langle T_{\tau} \Bigl\{ [
n_{\mu}^{c}(\mathbf{x}) \rho_{\sigma}(\mathbf{x})] [
n_{\nu}^{c}(\mathbf{x}') \rho_{\sigma'}(\mathbf{x}')] \Bigr\}
\Bigr\rangle_{c} , \nn \Pi_{\mu\nu}^{cc}(\mathbf{x},\mathbf{x'})
&=& t^{2} \Bigl\langle T_{\tau} \Bigl\{ [ J_{\sigma
\mu}(\mathbf{x})] [ J_{\sigma' \nu}(\mathbf{x}')] \Bigr\}
\Bigr\rangle_{c} - 4 i t \lambda_{so} \Bigl\langle T_{\tau}
\Bigl\{ [ n_{\mu}^{c}(\mathbf{x}) \sigma
\rho_{\sigma}(\mathbf{x})] [ J_{\sigma' \nu}(\mathbf{x}')] \Bigr\}
\Bigr\rangle_{c} - \Bigl( \mathbf{x} \leftrightarrow \mathbf{x}',
\mu \leftrightarrow \nu \Bigr) \nn &-& 16 \lambda_{so}^{2}
\Bigl\langle T_{\tau} \Bigl\{ [ n_{\mu}^{c}(\mathbf{x}) \sigma
\rho_{\sigma}(\mathbf{x})] [ n_{\nu}^{c}(\mathbf{x}') \sigma'
\rho_{\sigma'}(\mathbf{x}')] \Bigr\} \Bigr\rangle_{c} , \nn
\Pi_{\mu\nu}^{sc}(\mathbf{x},\mathbf{x'}) &=& t^{2} \Bigl\langle
T_{\tau} \Bigl\{ [ \sigma J_{\sigma \mu}(\mathbf{x})] [ J_{\sigma'
\nu}(\mathbf{x}')] \Bigr\} \Bigr\rangle_{c} - 4 i t \lambda_{so}
\Bigl\langle T_{\tau} \Bigl\{ [ n_{\mu}^{c}(\mathbf{x})
\rho_{\sigma}(\mathbf{x})] [ J_{\sigma' \nu}(\mathbf{x}')] \Bigr\}
\Bigr\rangle_{c} - \Bigl( \mathbf{x} \leftrightarrow \mathbf{x}',
\mu \leftrightarrow \nu \Bigr) \nn &-& 16 \lambda_{so}^{2}
\Bigl\langle T_{\tau} \Bigl\{ [ n_{\mu}^{c}(\mathbf{x})
\rho_{\sigma}(\mathbf{x})] [ n_{\nu}^{c}(\mathbf{x}') \sigma'
\rho_{\sigma'}(\mathbf{x}')] \Bigr\} \Bigr\rangle_{c} ,
\label{Correlation_Function_Def} \eqa
\end{widetext} where the summation for $\sigma = \pm$ should be
performed. $n_{\mu}^{c} = \frac{1}{2} z_{\alpha}^{c \dagger}
\sigma^{\mu}_{\alpha\beta} z_{\beta}^{c}$ represents the skyrmion
configuration of the spin direction $\mu$, and \bqa && J_{\sigma
\mu}(\mathbf{x}) = \psi_{\sigma}^{\dagger}(\mathbf{x})
[\partial_{\mu}\psi_{\sigma}(\mathbf{x})] - [\partial_{\mu}
\psi_{\sigma}^{\dagger}(\mathbf{x})] \psi_{\sigma}(\mathbf{x}) ,
\nn && \rho_{\sigma}(\mathbf{x}) =
\psi_{\sigma}^{\dagger}(\mathbf{x}) \psi_{\sigma}(\mathbf{x})
\label{Current_Density} \eqa denote the current and density with
spin $\sigma$. $\Pi_{\mu\nu}^{ss}(\mathbf{x},\mathbf{x'})$,
$\Pi_{\mu\nu}^{cc}(\mathbf{x},\mathbf{x'})$, and
$\Pi_{\mu\nu}^{sc}(\mathbf{x},\mathbf{x'})$ are the
spin-current--spin-current, charge-current--charge-current, and
spin-current--charge-current correlation functions, respectively.
The subscript $c$ represents ``connected."

Applying the Wick's theorem to the above expression, we obtain
\begin{widetext} \bqa \Pi_{\mu\nu}^{ss}(\mathbf{x},\mathbf{x'}) &=& -
t^{2} \sum_{\sigma = \pm} \Bigl[ \Bigl\{
\partial_{x_\mu} G_{\sigma\sigma}(\mathbf{x},\mathbf{x'}) \Bigr\} \Bigl\{
\partial_{x'_\nu} G_{\sigma\sigma}(\mathbf{x'},\mathbf{x}) \Bigr\}
- \Bigl\{
\partial_{x_\mu} \partial_{x'_\nu} G_{\sigma\sigma}(\mathbf{x},\mathbf{x'})
\Bigr\} G_{\sigma\sigma}(\mathbf{x'},\mathbf{x}) \nn &-&
G_{\sigma\sigma}(\mathbf{x},\mathbf{x'}) \Bigl\{
\partial_{x_\mu} \partial_{x'_\nu} G_{\sigma\sigma}(\mathbf{x'},\mathbf{x}) \Bigr\} + \Bigl\{
\partial_{x'_\nu} G_{\sigma\sigma}(\mathbf{x},\mathbf{x'}) \Bigr\} \Bigl\{
\partial_{x_\mu} G_{\sigma\sigma}(\mathbf{x'},\mathbf{x}) \Bigr\}
\Bigr] \nn &+& 4 i t \lambda_{so} n_{\mu}^{c}(\boldsymbol{x})
\sum_{\sigma = \pm} \sigma \Bigl[ G_{\sigma\sigma}(
\mathbf{x},\mathbf{x}') \Bigl\{
\partial_{x_{\nu}'} G_{\sigma\sigma}(\mathbf{x}',\mathbf{x})
\Bigr\} - \Bigl\{ \partial_{x_{\nu}'}
G_{\sigma\sigma}(\mathbf{x},\mathbf{x}') \Bigr\}
G_{\sigma\sigma}(\mathbf{x}',\mathbf{x}) \Bigr] + \Bigl(
\mathbf{x} \leftrightarrow \mathbf{x}', \mu \leftrightarrow \nu
\Bigr) \nn &+& 16 \lambda_{so}^{2} n_{\mu}^{c}(\boldsymbol{x})
n_{\nu}^{c}(\boldsymbol{x}') \sum_{\sigma = \pm}
G_{\sigma\sigma}(\boldsymbol{x},\boldsymbol{x}')
G_{\sigma\sigma}(\boldsymbol{x}',\boldsymbol{x}) , \nn
\Pi_{\mu\nu}^{cc}(\mathbf{x},\mathbf{x'}) &=&
\Pi_{\mu\nu}^{ss}(\mathbf{x},\mathbf{x'}) , \nn
\Pi_{\mu\nu}^{sc}(\mathbf{x},\mathbf{x'}) &=& - t^{2} \sum_{\sigma
= \pm} \sigma \Bigl[ \Bigl\{
\partial_{x_\mu} G_{\sigma\sigma}(\mathbf{x},\mathbf{x'}) \Bigr\} \Bigl\{
\partial_{x'_\nu} G_{\sigma\sigma}(\mathbf{x'},\mathbf{x}) \Bigr\}
+ \Bigl\{
\partial_{x_\mu} \partial_{x'_\nu} G_{\sigma\sigma}(\mathbf{x},\mathbf{x'})
\Bigr\} G_{\sigma\sigma}(\mathbf{x'},\mathbf{x}) \nn &+&
G_{\sigma\sigma}(\mathbf{x},\mathbf{x'}) \Bigl\{
\partial_{x_\mu} \partial_{x'_\nu} G_{\sigma\sigma}(\mathbf{x'},\mathbf{x})
\Bigr\} - \Bigl\{
\partial_{x'_\nu} G_{\sigma\sigma}(\mathbf{x},\mathbf{x'}) \Bigr\} \Bigl\{
\partial_{x_\mu} G_{\sigma\sigma}(\mathbf{x'},\mathbf{x}) \Bigr\} \Bigr]
\nn &+& 4 i t \lambda_{so} n_{\mu}^{c}(\boldsymbol{x})
\sum_{\sigma = \pm} \Bigl[ G_{\sigma\sigma}(
\mathbf{x},\mathbf{x}') \Bigl\{
\partial_{x_{\nu}'} G_{\sigma\sigma}(\mathbf{x}',\mathbf{x})
\Bigr\} - \Bigl\{ \partial_{x_{\nu}'}
G_{\sigma\sigma}(\mathbf{x},\mathbf{x}') \Bigr\}
G_{\sigma\sigma}(\mathbf{x}',\mathbf{x}) \Bigr] + \Bigl(
\mathbf{x} \leftrightarrow \mathbf{x}', \mu \leftrightarrow \nu
\Bigr) \nn &+& 16 \lambda_{so}^{2} n_{\mu}^{c}(\boldsymbol{x})
n_{\nu}^{c}(\boldsymbol{x}') \sum_{\sigma = \pm} \sigma
G_{\sigma\sigma}(\boldsymbol{x},\boldsymbol{x}')
G_{\sigma\sigma}(\boldsymbol{x}',\boldsymbol{x}) ,
\label{Correlation_Green_FT} \eqa
\end{widetext} where the single
particle Green's function is given by \bqa &&
G_{\sigma\sigma}(\mathbf{x},\mathbf{x'}) = - \Bigl\langle T_{\tau}
\Bigl\{ \psi_{\sigma}(\mathbf{x})
\psi_{\sigma}^{\dagger}(\mathbf{x'})  \Bigr\} \Bigr\rangle .
\label{Green_FT} \eqa

\subsection{Single particle Green's function}

It is an essential procedure to find the single particle Green's
function in the presence of the single skyrmion potential, where
the plane wave basis cannot be applied due to translational
symmetry breaking. An important feature that we try to catch is
time reversal symmetry breaking to allow Hall conductivity, where
an internal magnetic flux of the skyrmion gives rise to the Hall
motion of  itinerant electrons. This effect will be introduced,
solving the Schrodinger equation in the single skyrmion potential
and finding new eigen states.

We start from the Schrodinger equation with the single skyrmion
potential in the polar coordinate \begin{widetext} \bqa && ( - i
\partial_{t} - \mu_{r} - J_{H} S \sigma ) \psi_{\sigma}(r, \phi,
t) \nn && - t \Bigl\{ \Bigl( \frac{\partial}{\partial r} - i
\sigma \frac{2\lambda_{so}}{t a} \frac{2 \xi r }{\xi^{2} + r^{2} }
\Bigr)^{2} + \frac{1}{r}\Bigl( \frac{\partial}{\partial r} - i
\sigma \frac{2\lambda_{so}}{t a} \frac{2 \xi r }{\xi^{2} + r^{2} }
\Bigr) + \frac{1}{r^{2}} \Bigl(
\partial_{\phi} + i \sigma \frac{ r^{2} }{\xi^{2} + r^{2}}
\Bigr)^{2} \Bigr\}  \psi_{\sigma}(r, \phi, t) = 0 ,
\label{Shrodinger_Eq} \eqa
%
%
\end{widetext}
where $\xi$ is the skyrmion core size. In appendix B1 we show its
derivation.

A standard way to solve this kind of equation is to expand the
wave function in a complete basis \bqa && \psi_{\sigma}(r, \phi,
t) = \int d E_{\sigma} e^{- i E_{\sigma} t} \sum_{n}
a_{\sigma}(n,E_{\sigma}) \Psi_{n}^{\sigma}(r,\phi;E_{\sigma}) ,
\nn \label{Expansion} \eqa where $E_{\sigma}$ is an energy eigen
value and $n$ is a good quantum number.
$\Psi_{n}^{\sigma}(r,\phi;E_{\sigma})$ is an eigen state described
by two conserving quantum numbers, $n$ and $E_{\sigma}$, and
$a_{\sigma}(n,E_{\sigma})$ is an associated coefficient in the
expansion, identified with an annihilation operator in the second
quantization expression. The rotational symmetry of the skyrmion
potential leads us to the following decomposition \bqa &&
\Psi_{n}^{\sigma}(r,\phi;E_{\sigma}) = \mathcal{C} e^{in\phi}
F_{\sigma}^{n}(r;E_{\sigma}) , \label{Ansatz} \eqa where $n$ is
identified with an angular momentum and
$F_{\sigma}^{n}(r;E_{\sigma})$ is the corresponding radial wave
function. $\mathcal{C}$ is the normalization constant, determined
by \bqa && \int_{0}^{2\pi} d \phi \int_{0}^{\infty} d r r \sum_{n}
\Psi_{n}^{\sigma *}(r,\phi;E_{\sigma})
\Psi_{n}^{\sigma}(r,\phi;E_{\sigma}\blue{'}) \nn && =
\delta(E_{\sigma}-E_{\sigma}') . \eqa

Unfortunately, we fail to find an analytic expression for the
radial wave function in the presence of the spin-orbit
interaction. Instead, we could obtain the most general expression
for the radial wave function in the absence of the spin-orbit
coupling
\begin{widetext} \bqa F_{\sigma}^{n}(r) &=& \mathcal{C}_{1} (
\xi^{2}+r^{2})^{\frac{1+\sqrt{2}}{2}} r^{n}
\mbox{HeunC}\Bigl(0,n,\sqrt{2},\frac{1+n}{2}-\frac{\xi^{2}}{4 }
\mathcal{E}_{\sigma}, \frac{1- \sigma n}{2} + \frac{\xi^{2}}{4 }
\mathcal{E}_{\sigma}, - \frac{r^{2}}{\xi^{2}} \Bigr) \nn &+&
\mathcal{C}_{2} ( \xi^{2}+r^{2})^{\frac{1+\sqrt{2}}{2}} r^{-n}
\mbox{HeunC}\Bigl(0,-n,\sqrt{2},\frac{1-n}{2}-\frac{\xi^{2}}{4 }
\mathcal{E}_{\sigma}, \frac{1- \sigma n}{2} + \frac{\xi^{2}}{4 }
\mathcal{E}_{\sigma}, - \frac{r^{2}}{\xi^{2}} \Bigr) ,
\label{Solution} \eqa
\end{widetext}
where
$\mbox{HeunC}\Bigl(0,n,\sqrt{2},\frac{1+n}{2}-\frac{\xi^{2}}{4 }
\mathcal{E}_{\sigma}, \frac{1- \sigma n}{2} + \frac{\xi^{2}}{4 }
\mathcal{E}_{\sigma}, - \frac{r^{2}}{\xi^{2}} \Bigr)$ is a
solution of the Heun's confluent equation, regarded as the
generalization of the hypergeometric function when an ordinary
differential equation contains four singularities \cite{Heun}.
$\mathcal{E}_{\sigma} \equiv t^{-1} (E_{\sigma} + \mu_{r} + J_{H}
S \sigma)$ is an effective energy level. In appendix B2 we discuss
the Heun's confluent equation in detail.

The main feature of this wave function lies in the time reversal
symmetry breaking. An effective magnetic flux due to the skyrmion
gives rise to an effective Lorentz force for itinerant electrons,
resulting in chirality to the electron state. One can see this
effect from the difference between two time reversal states with
opposite angular momenta.
%
%
In the asymptotic limit of $r \rightarrow \infty$ the Heun's
confluent function becomes
\begin{widetext} \bqa F_{\sigma}^{n}(r) = \mathcal{C}'_{1} (
\xi^{2}+r^{2})^{\frac{1+\sqrt{2}}{2}} r^{n} \exp\Bigl( -
\frac{\mathcal{E}_{\sigma}}{4(n + 2 + \sqrt{2})} r^{2} \Bigr) +
\mathcal{C}'_{2} ( \xi^{2}+r^{2})^{\frac{1+\sqrt{2}}{2}} r^{-n}
\exp\Bigl( - \frac{\mathcal{E}_{\sigma}}{4(- n + 2 + \sqrt{2})}
r^{2} \Bigr) , \nonumber \eqa \end{widetext} as shown in appendix
B2. This asymptotic wave function reveals that quantum states with
negative angular momenta are suppressed due to the effective
Lorentz force, imposing the condition that the wave function
should not diverge in the $r \rightarrow \infty$ limit. If we
consider positive angular momenta in the above expression, the
second term allows only $n = 1, 2$ due to the regularity
condition, thus forcing $\mathcal{C}'_{2}$ to vanish identically.
On the other hand, the first term allows only $n = -1, -2$ among
negative angular momenta, giving rise to $\mathcal{C}'_{1} = 0$.
Considering that either case reaches the same expression, the
consistent solution is given by only the first term with positive
angular momenta.

It is straightforward to find the single particle Green's function
in the new basis. Inserting Eq. (\ref{Expansion}) with Eq.
(\ref{Ansatz}) into the Green's function [Eq. (\ref{Green_FT})]
and performing the Fourier transformation, we obtain
\begin{widetext} \bqa && G_{\sigma\sigma}(r,r',\phi-\phi',i\omega) =
|\mathcal{C}|^{2} \int d \mathcal{E}_{\sigma} \sum_{n}
e^{in(\phi-\phi')} \frac{F_{\sigma}^{n}(r;\mathcal{E}_{\sigma})
F_{\sigma}^{n}(r';\mathcal{E}_{\sigma})}{i\omega -
\Sigma_{\sigma}(i\omega,\mathcal{E}_{\sigma}) -
\mathcal{E}_{\sigma} } , \label{Green_FT_Final} \eqa
\end{widetext} where the self-energy correction \bqa &&
\Sigma_{\sigma}(i\omega,\mathcal{E}_{\sigma}) =
\Sigma_{\sigma}^{imp}(i\omega,\mathcal{E}_{\sigma}) +
\Sigma_{\sigma}^{ele}(i\omega,\mathcal{E}_{\sigma}) \nonumber \eqa
is introduced, resulting from both elastic impurity scattering and
inelastic interaction effects with gauge fluctuations,
respectively. A way to understand this expression is shown in
appendix B3.

\subsection{Longitudinal conductivity}

Inserting the Green's function [Eq. (\ref{Green_FT_Final})] into
the conductivity tensor [Eq. (\ref{Correlation_Green_FT})] with
the following representation \bqa &&
\partial_{x} = \cos \phi
\partial_{r} - \frac{\sin \phi}{r} \partial_{\phi} , ~~~~~
\partial_{y} = \sin \phi
\partial_{r} + \frac{\cos \phi}{r}
\partial_{\phi} \nn \eqa in the polar coordinate, we are ready to calculate
the longitudinal ``spin conductivity"
\begin{widetext}
\bqa && \sigma_{sp}^{\psi} = - \int_{0}^{\infty} dr r
\int_{0}^{\infty} dr' r' \int_{0}^{2\pi} d \phi \int_{0}^{2\pi} d
\phi' \lim_{\Omega \rightarrow 0} \frac{\Im
\Pi_{xx}^{ss}(r,r',\phi-\phi',\Omega+i\delta)}{\Omega} ,
\label{Longitudinal_Conductivity} \eqa
\end{widetext}

A lengthy but straightforward calculation, which is given in
appendix C1, yields
\begin{widetext}
\bqa \sigma_{sp}^{\psi}
&=& 2 \pi^{2} t^{2} |\mathcal{C}|^{4} \sum_{\sigma = \pm} \int d
\mathcal{E}_{\sigma} \int d \mathcal{E}_{\sigma}'
\int_{-\infty}^{\infty} d \nu_{\sigma} \Bigl( - \frac{\partial
f(\Omega)}{\partial \Omega} \Bigr)_{\Omega = \nu_{\sigma}}
{\cal A}_{\sigma}(\nu_{\sigma},\mathcal{E}_{\sigma}) {\cal
A}_{\sigma}(\nu_{\sigma},\mathcal{E}_{\sigma}') \sum_{n}
[\mathcal{R}_{\sigma}^n(\mathcal{E}_{\sigma};\mathcal{E}_{\sigma}')]^{2}
\nn &+& 16 \pi^{2} \lambda_{so}^{2} |\mathcal{C}|^{4} \sum_{\sigma
= \pm} \int d \mathcal{E}_{\sigma} \int d \mathcal{E}_{\sigma}'
\int_{-\infty}^{\infty} d \nu_{\sigma} \Bigl( - \frac{\partial
f(\Omega)}{\partial \Omega} \Bigr)_{\Omega = \nu_{\sigma}} {\cal
A}_{\sigma}(\nu_{\sigma},\mathcal{E}_{\sigma}) {\cal
A}_{\sigma}(\nu_{\sigma},\mathcal{E}_{\sigma}') \sum_{n}
[\mathcal{S}_{\sigma}^n(\mathcal{E}_{\sigma};\mathcal{E}_{\sigma}')]^{2}
\label{Sigma_XX} , \eqa
\end{widetext}
where ${\cal A}_{\sigma}(\nu_{\sigma},\mathcal{E}_{\sigma})$ is
the spectral function for  itinerant electrons,
\begin{widetext} \bqa &&
{\cal A}_{\sigma}(\nu_{\sigma},\mathcal{E}_{\sigma}) = -
\frac{1}{\pi} \frac{\Im
\Sigma_{\sigma}(\nu_{\sigma,\mathcal{E}_{\sigma}})}{[\nu_{\sigma}
- \Re \Sigma(\nu_{\sigma},\mathcal{E}_{\sigma}) -
\mathcal{E}_{\sigma}]^{2} + [\Im
\Sigma_{\sigma}(\nu_{\sigma},\mathcal{E}_{\sigma})]^{2}} ,
\label{Spectral_FT} \eqa and
$\mathcal{R}_{\sigma}^n(\mathcal{E}_{\sigma};\mathcal{E}_{\sigma}')$
and
$\mathcal{S}_{\sigma}^n(\mathcal{E}_{\sigma};\mathcal{E}_{\sigma}')$
are composed of radial wave functions, given by \bqa &&
\mathcal{R}_{\sigma}^n(\mathcal{E}_{\sigma};\mathcal{E}_{\sigma}')
= \int_{0}^{\infty} dr r \Bigl( [\partial_{r}
F_{\sigma}^{n}(r;\mathcal{E}_{\sigma})]
F_{\sigma}^{n+1}(r;\mathcal{E}_{\sigma}') -
F_{\sigma}^{n}(r;\mathcal{E}_{\sigma}) [\partial_{r}
F_{\sigma}^{n+1}(r;\mathcal{E}_{\sigma}') ] \Bigr)  , \nn &&
\mathcal{S}_{\sigma}^n(\mathcal{E}_{\sigma};\mathcal{E}_{\sigma}')
= \int_{0}^{\infty} dr r f(r)
F_{\sigma}^{n}(r;\mathcal{E}_{\sigma})
F_{\sigma}^{n+1}(r;\mathcal{E}'_{\sigma}) , \label{R_FT} \eqa
\end{widetext}
respectively. $f(r) = \frac{2 \xi r}{\xi^{2} + r^{2}}$ in
$\mathcal{S}_{\sigma}^n(\mathcal{E}_{\sigma};\mathcal{E}_{\sigma}')$
originates from the skyrmion configuration of the spin component.
We note that only nearest neighbor angular-momentum channels are
coupled. The cross correlation between the spin-current and
spin-density, proportional to $t \lambda_{so}$ in Eq.
(\ref{Correlation_Function_Def}), vanishes for the longitudinal
conductivity.

If we consider the non-interacting limit of
$\Re\Sigma_{\sigma}(\nu_{\sigma}),
\Im\Sigma_{\sigma}(\nu_{\sigma}) \rightarrow 0$, we can simplify
this expression further, resorting to the asymptotic form of the
radial wave function in the case of $\lambda_{so} = 0$. This
procedure is shown in appendix C2, and the final analytic
expression is given by
\begin{widetext} \bqa \sigma_{sp}^{\psi}(T \rightarrow 0) &=& 2
\pi^{2} t^{2} |\mathcal{C}|^{4} (\mathcal{C}'_{1})^{4}
\xi^{2(1+\sqrt{2})} \sum_{n} L^{2n+2} \Bigl\{ \frac{{}_2F_1[1 + n,
-\sqrt{2}, 2 + n, -(L/\xi)^2]}{2(n+1)} \nn &+& \Bigl(
\frac{L}{\xi} \Bigr)^{2} \frac{{}_2F_1[2 + n, -\sqrt{2}, 3 + n,
-(L/\xi)^2]}{2(n+2)} \Bigr\} , \label{Asymptotic_Sigma_XX} \eqa
\end{widetext} where $L$ is the system size and $\xi$ is the
skyrmion core size. The hypergeometric function is defined in the
Mathematica program.

As discussed before, the electrical conductivity is the same as
the spin conductivity \bqa && \sigma_{el}^{\psi} =
\sigma_{sp}^{\psi} . \eqa The cross longitudinal conductivity is
given by \begin{widetext} \bqa \sigma_{sp-el}^{\psi} &=& 2 \pi^{2}
t^{2} |\mathcal{C}|^{4} \sum_{\sigma= \pm} \sigma \int d
\mathcal{E}_{\sigma} \int d \mathcal{E}_{\sigma}'
\int_{-\infty}^{\infty} d \nu_{\sigma} \Bigl( - \frac{\partial
f(\Omega)}{\partial \Omega} \Bigr)_{\Omega = \nu_{\sigma}} {\cal
A}_{\sigma}(\nu_{\sigma},\mathcal{E}_{\sigma}) {\cal
A}_{\sigma}(\nu_{\sigma},\mathcal{E}_{\sigma}') \sum_{n}
[\mathcal{R}_{\sigma}^n(\mathcal{E}_{\sigma};\mathcal{E}_{\sigma}')]^{2}
\nn &+& 16 \pi^{2} \lambda_{so}^{2} |\mathcal{C}|^{4} \sum_{\sigma
= \pm} \sigma \int d \mathcal{E}_{\sigma} \int d
\mathcal{E}_{\sigma}' \int_{-\infty}^{\infty} d \nu_{\sigma}
\Bigl( - \frac{\partial f(\Omega)}{\partial \Omega} \Bigr)_{\Omega
= \nu_{\sigma}} {\cal
A}_{\sigma}(\nu_{\sigma},\mathcal{E}_{\sigma}) {\cal
A}_{\sigma}(\nu_{\sigma},\mathcal{E}_{\sigma}') \sum_{n}
[\mathcal{S}_{\sigma}^n(\mathcal{E}_{\sigma};\mathcal{E}_{\sigma}')]^{2}
. \eqa
\end{widetext} As mentioned before, this correlation function is finite
because we are considering ferromagnetism. But, this cross effect
is not crucial for the skyrmion dynamics, just modifying the
transport coefficient.

\subsection{Hall conductivity}

The Hall conductivity can be obtained along the same strategy as
the longitudinal conductivity.
%
A straightforward calculation given in appendix D1 leads to the
following expression,
\begin{widetext}
\bqa &&
\sigma_{sH}^{\psi}
= 2 \pi^{2} t^{2} |\mathcal{C}|^{4} \sum_{\sigma = \pm} \int d
\mathcal{E}_{\sigma} \int d \mathcal{E}'_{\sigma}
\int_{-\infty}^{\infty} d \nu_{\sigma} \int_{-\infty}^{\infty} d
\nu_{\sigma}' \frac{f(\nu_{\sigma}) - f(\nu_{\sigma}')}{
(\nu_{\sigma} -\nu_{\sigma}')} {\cal
A}_{\sigma}(\nu_{\sigma},\mathcal{E}_{\sigma}) {\cal
A}_{\sigma}(\nu_{\sigma}',\mathcal{E}_{\sigma}') \nn &&
\times\sum_{n} \Bigl\{
\mathcal{O}_{\sigma}^n(\mathcal{E}_{\sigma};\mathcal{E}_{\sigma}')
\mathcal{R}_{\sigma}^n(\mathcal{E}_{\sigma};\mathcal{E}_{\sigma}')
-
\mathcal{P}_{\sigma}^n(\mathcal{E}_{\sigma};\mathcal{E}_{\sigma}')
\mathcal{Q}_{\sigma}^n(\mathcal{E}_{\sigma};\mathcal{E}_{\sigma}')
\Bigr\} \nn && + 8 \pi^{2} t \lambda_{so} |\mathcal{C}|^{4}
\sum_{\sigma = \pm} \sigma \int d \mathcal{E}_{\sigma} \int d
\mathcal{E}'_{\sigma} \int_{-\infty}^{\infty} d \nu_{\sigma}
\int_{-\infty}^{\infty} d \nu_{\sigma}' \frac{f(\nu_{\sigma}) -
f(\nu_{\sigma}')}{ (\nu_{\sigma} -\nu_{\sigma}')} {\cal
A}_{\sigma}(\nu_{\sigma},\mathcal{E}_{\sigma}) {\cal
A}_{\sigma}(\nu_{\sigma}',\mathcal{E}_{\sigma}') \sum_{n}
\mathcal{O}_{\sigma}^n(\mathcal{E}_{\sigma};\mathcal{E}_{\sigma}')
\mathcal{S}_{\sigma}^n(\mathcal{E}_{\sigma};\mathcal{E}_{\sigma}')
\nn && + 16 \pi^{2} \lambda_{so}^{2} |\mathcal{C}|^{4}
\sum_{\sigma = \pm} \int d \mathcal{E}_{\sigma} \int d
\mathcal{E}'_{\sigma} \int_{-\infty}^{\infty} d \nu_{\sigma}
\int_{-\infty}^{\infty} d \nu_{\sigma}' \frac{f(\nu_{\sigma}) -
f(\nu_{\sigma}')}{ (\nu_{\sigma} -\nu_{\sigma}')} {\cal
A}_{\sigma}(\nu_{\sigma},\mathcal{E}_{\sigma}) {\cal
A}_{\sigma}(\nu_{\sigma}',\mathcal{E}_{\sigma}') \sum_{n} [
\mathcal{S}_{\sigma}^n(\mathcal{E}_{\sigma};\mathcal{E}_{\sigma}')
]^{2} , \label{Sigma_XY} \eqa where \bqa &&
\mathcal{O}_{\sigma}^n(\mathcal{E}_{\sigma};\mathcal{E}_{\sigma}')
= \int_{0}^{\infty} dr F_{\sigma}^{n}(r;\mathcal{E}_{\sigma})
F_{\sigma}^{n+1}(r;\mathcal{E}'_{\sigma}) , ~~~~~
\mathcal{P}_{\sigma}^n(\mathcal{E}_{\sigma};\mathcal{E}_{\sigma}')
= \int_{0}^{\infty} dr r F_{\sigma}^{n}(r;\mathcal{E}_{\sigma})
F_{\sigma}^{n+1}(r;\mathcal{E}'_{\sigma}) , \nn &&
\mathcal{Q}_{\sigma}^n(\mathcal{E}_{\sigma};\mathcal{E}_{\sigma}')
= \int_{0}^{\infty} dr \Bigl( [\partial_{r}
F_{\sigma}^{n}(r;\mathcal{E}_{\sigma})]
F_{\sigma}^{n+1}(r;\mathcal{E}_{\sigma}') -
F_{\sigma}^{n}(r;\mathcal{E}_{\sigma}) [\partial_{r}
F_{\sigma}^{n+1}(r;\mathcal{E}_{\sigma}') ] \Bigr) , \nn &&
\mathcal{R}_{\sigma}^n(\mathcal{E}_{\sigma};\mathcal{E}_{\sigma}')
= \int_{0}^{\infty} dr r \Bigl( [\partial_{r}
F_{\sigma}^{n}(r;\mathcal{E}_{\sigma})]
F_{\sigma}^{n+1}(r;\mathcal{E}_{\sigma}') -
F_{\sigma}^{n}(r;\mathcal{E}_{\sigma}) [\partial_{r'}
F_{\sigma}^{n+1}(r;\mathcal{E}_{\sigma}') ] \Bigr) , \nn &&
\mathcal{S}_{\sigma}^n(\mathcal{E}_{\sigma};\mathcal{E}_{\sigma}')
= \int_{0}^{\infty} dr r f(r)
F_{\sigma}^{n}(r;\mathcal{E}_{\sigma})
F_{\sigma}^{n+1}(r;\mathcal{E}'_{\sigma}) . \label{FTs_O_P_Q_R_S}
\eqa
\end{widetext}
The spin-orbit interaction contributes to the Hall conductivity.

This expression reveals that the Hall conductivity does not vanish
even without the spin-orbit coupling because the $r$ integration
differs from the $r'$ integration. The underlying mechanism is
that the chirality is preferred, reflected in the Heun's confluent
function. As a result, the $r$ integral becomes different from the
$r'$ integral, originating from the angular dependence.

The finite Hall conductivity becomes clearer if one focuses on the
asymptotic expression of the radial wave function. We simplify Eq.
(\ref{Sigma_XY}) further in the non-interaction limit, shown in
appendix D2. We note that the exponential decay of the asymptotic
form is expected to make the contribution from the radial integral
finite for the Hall conductivity. Since this spin Hall
conductivity corresponds to $\sigma_{ss}^{H} = \int d x \int d y
\Theta_{ss}(\boldsymbol{x})$, the finite Hall conductivity means
that it is not proportional to $L^{2}$ as discussed before.


The charge Hall conductivity is the same as the spin Hall
conductivity \bqa && \sigma_{eH}^{\psi} = \sigma_{sH}^{\psi} ,
\eqa implying that $\sigma_{cc}^{H}$ is a constant, not
proportional to $L^{2}$ as discussed in Eq. (\ref{SkV_Insulator}).

The cross Hall coefficient corresponding to $\sigma_{sc}^{H}$ in
Eq. (\ref{Hall_Conductances}) is given by
\begin{widetext}
\bqa && \sigma_{seH}^{\psi} = 2 \pi^{2} t^{2} |\mathcal{C}|^{4}
\sum_{\sigma = \pm} \sigma \int d \mathcal{E}_{\sigma} \int d
\mathcal{E}'_{\sigma} \int_{-\infty}^{\infty} d \nu_{\sigma}
\int_{-\infty}^{\infty} d \nu_{\sigma}' \frac{f(\nu_{\sigma}) -
f(\nu_{\sigma}')}{ (\nu_{\sigma} -\nu_{\sigma}')} {\cal
A}_{\sigma}(\nu_{\sigma},\mathcal{E}_{\sigma}) {\cal
A}_{\sigma}(\nu_{\sigma}',\mathcal{E}_{\sigma}') \nn &&
\times\sum_{n} \Bigl\{
\mathcal{O}_{\sigma}^n(\mathcal{E}_{\sigma};\mathcal{E}_{\sigma}')
\mathcal{R}_{\sigma}^n(\mathcal{E}_{\sigma};\mathcal{E}_{\sigma}')
-
\mathcal{P}_{\sigma}^n(\mathcal{E}_{\sigma};\mathcal{E}_{\sigma}')
\mathcal{Q}_{\sigma}^n(\mathcal{E}_{\sigma};\mathcal{E}_{\sigma}')
\Bigr\} \nn && + 8 \pi^{2} t \lambda_{so} |\mathcal{C}|^{4}
\sum_{\sigma = \pm} \int d \mathcal{E}_{\sigma} \int d
\mathcal{E}'_{\sigma} \int_{-\infty}^{\infty} d \nu_{\sigma}
\int_{-\infty}^{\infty} d \nu_{\sigma}' \frac{f(\nu_{\sigma}) -
f(\nu_{\sigma}')}{ (\nu_{\sigma} -\nu_{\sigma}')} {\cal
A}_{\sigma}(\nu_{\sigma},\mathcal{E}_{\sigma}) {\cal
A}_{\sigma}(\nu_{\sigma}',\mathcal{E}_{\sigma}') \sum_{n}
\mathcal{O}_{\sigma}^n(\mathcal{E}_{\sigma};\mathcal{E}_{\sigma}')
\mathcal{S}_{\sigma}^n(\mathcal{E}_{\sigma};\mathcal{E}_{\sigma}')
\nn && + 16 \pi^{2} \lambda_{so}^{2} |\mathcal{C}|^{4}
\sum_{\sigma = \pm} \sigma \int d \mathcal{E}_{\sigma} \int d
\mathcal{E}'_{\sigma} \int_{-\infty}^{\infty} d \nu_{\sigma}
\int_{-\infty}^{\infty} d \nu_{\sigma}' \frac{f(\nu_{\sigma}) -
f(\nu_{\sigma}')}{ (\nu_{\sigma} -\nu_{\sigma}')} {\cal
A}_{\sigma}(\nu_{\sigma},\mathcal{E}_{\sigma}) {\cal
A}_{\sigma}(\nu_{\sigma}',\mathcal{E}_{\sigma}') \sum_{n} [
\mathcal{S}_{\sigma}^n(\mathcal{E}_{\sigma};\mathcal{E}_{\sigma}')
]^{2} . \eqa
\end{widetext} This is
also non-vanishing because of the $J_{H} S$ term in the effective
action, resulting from the different population between $\uparrow$
and $\downarrow$ itinerant electrons, proportional to $J_{H} S$.

\section{Summary}

We investigated dynamics of skyrmions under spin currents driven
by electric field in itinerant ferromagnets. We developed a novel
framework based on the effective U(1) gauge theory formulation
[Eq. (\ref{EFT}) and Eq. (\ref{EFT_CS})], where the
Maxwell-Chern-Simons equation [Eq. (\ref{Maxwell_CS}) with Eq.
(\ref{Constituent_Eq})] is the key equation for soliton dynamics.
Although this framework differs from the Landau-Lifshitz-Gilbert
equation approach, both equations turn out to have essentially the
same ingredient as it should be. Indeed, we recovered the expected
result of the Landau-Lifshitz-Gilbert equation approach [Eq.
(\ref{LLG})] from the Maxwell equation framework [Eq.
(\ref{Skyrmion_Motion})].

An important improvement beyond the previous study lies in the
mutual feedback effect for both skyrmion dynamics and electron
motion, where the internal magnetic flux of the skyrmion gives
rise to the Hall motion to  itinerant electrons or vice versa.
This physics can be described by the Chern-Simons terms [Eq.
(\ref{EFT_CS})] in the skyrmion moving frame. As a result, we
revealed that the skyrmion motion follows not only the electric
field but also its transverse direction [Eq.
(\ref{Skyrmion_Velocity})]. An interesting observation is that
even if an insulating state is considered, the electric field will
induce the dissipationless skyrmion Hall current due to the
Chern-Simons terms [Eq. (\ref{SkV_Insulator})]. In particular, we
predict that the dissipationless skyrmion Hall current will
survive even in the thermodynamic limit as far as the spin-orbit
interaction is introduced, expected to realize in the surface
state of three dimensional topological insulators when magnetic
impurities are deposited.

We evaluated the Hall conductivity in the presence of the single
skyrmion potential, showing that it is non-vanishing indeed
although it will vanish in the thermodynamic limit because we are
considering only one skyrmion without the spin-orbit interaction.
A careful treatment is required because we should construct new
eigen basis in the single skyrmion system [Section IV-B]. Based on
this construction, we calculated all kinds of correlation
functions explicitly, and found the general expression for the
Hall conductivity [Eq. (\ref{Sigma_XY}) with Eq.
(\ref{FTs_O_P_Q_R_S})].

K.-S. Kim would like to thank H.-W. Lee for insightful
discussions, and K.-Y. Kim for helping us solve the Schrodinger
equation in the presence of the soliton background. K.-S. Kim was
supported by the National Research Foundation of Korea (NRF) grant
funded by the Korea government (MEST) (No. 2010-0074542). S. O.
was partly supported by Grants-in-Aid for Scientific Research
under Grant No. 19052006 from the MEXT of Japan and No. 21740275
from Japan Society of the Promotion of Science.

\appendix

\section{Mapping from the double exchange model to an effective U(1) gauge theory in the strong coupling limit}

We start from the ferromagnetic Kondo lattice model or the double
exchange model with the Rashba spin-orbit coupling \bqa && Z =
\int D c_{i\sigma} D \boldsymbol{\vec{S}}_{i} \exp\Bigl( - S_{B} -
\int_{0}^{\beta} d \tau L \Bigr) , \nn &&  L = \sum_{i}
c_{i\sigma}^{\dagger} (\partial_{\tau} - \mu) c_{i\sigma} - t
\sum_{ij} (c_{i\sigma}^{\dagger}c_{j\sigma} + H.c.) \nn && - J_{H}
\sum_{i}
c_{i\alpha}^{\dagger}(\boldsymbol{\vec{\sigma}}\cdot\boldsymbol{\vec{S}}_{i})_{\alpha\beta}c_{i\beta}
\nn && - i \lambda_{so} \sum_{i} \sigma^{\mathbf{a}}_{\alpha\beta}
(c_{i+\mathbf{\hat{a}} \alpha}^{\dagger} c_{i \beta} - c_{i
\alpha}^{\dagger} c_{i+\mathbf{\hat{a}} \beta}) , \eqa where
$c_{i\sigma}$ represents the conduction electron field and
$\boldsymbol{\vec{S}}_{i}$ expresses the localized spin. $\mu$ is
an electron chemical potential and $t$ is the wave-function
overlap integral for conduction electrons. $J_{H}$ is the exchange
coupling constant, set to be positive. $\lambda_{so}$ is the
Rashba spin-orbit coupling constant, which breaks the inversion
symmetry, realized on the surface or in the inversion-symmetry
breaking material. $S_{B}$ is the single-spin Berry phase term in
the spin coherent-state representation, given by \bqa && S_{B} = i
S \int_{0}^{\beta} d \tau \sum_{i}
\partial_{\tau} \phi_{i} (1 - \cos \theta_{i}) , \eqa where the
spin field is expressed by two angles, $\boldsymbol{\vec{S}}_{i} =
(\sin\theta_{i}\cos\phi_{i},\sin\theta_{i}\sin\phi_{i},\cos\theta_{i})$.

If we consider the system such that dynamics of localized spins is
much slower than that of itinerant electrons, spins of conduction
electrons follow localized spins. This situation allows us to take
the strong coupling approach \bqa - J_{H}
\boldsymbol{\vec{S}_{i}}\cdot\boldsymbol{\vec{\sigma}}_{\alpha\beta}
&=& - J_{H} S U_{i\alpha\gamma} \sigma^{z}_{\gamma\delta}
U_{i\delta\beta}^{\dagger} , \nn \psi_{i\alpha} &=&
U_{i\alpha\beta}^{\dagger}c_{i\beta} ,  \eqa where the unitary
matrix field $\boldsymbol{U}_{i} =
\left( \begin{array}{cc} z_{i\uparrow} & z_{i\downarrow}^{\dagger} \\
z_{i\downarrow} & - z_{i\uparrow}^{\dagger} \end{array} \right)$
consists of a bosonic spinon field $z_{i\sigma}$ and the alignment
of the itinerant spin to the localized spin introduces a
 electron field $\psi_{i\alpha}$. The spinon field can
be represented in the following way, $z_{i\uparrow} = e^{-i
\frac{\phi_{i}}{2}} \cos \frac{\theta_{i}}{2}$ and
$z_{i\downarrow} = e^{i \frac{\phi_{i}}{2}} \sin
\frac{\theta_{i}}{2}$, which gives the self-consistent expression
$\boldsymbol{\vec{S}}_{i} = \frac{1}{2} z_{i\alpha}^{\dagger}
\boldsymbol{\vec{\sigma}}_{\alpha\beta} z_{i\beta}$.

Representing Eq. (A1) for $c_{i\sigma}$ and
$\boldsymbol{\vec{S}}_{i}$ in terms of $\psi_{i\alpha}$ and
$U_{i\alpha\beta}$ [Eq. (A3)],
%
%
and taking the continuum approximation, we reach the following
expression \begin{widetext} \bqa && Z = \int D \psi_{\alpha} D
U_{\alpha\beta} \delta(U^{\dagger}_{\alpha\gamma}U_{\gamma\beta} -
\delta_{\alpha\beta}) e^{ - S_{B} - \int_{0}^{\beta} d \tau \int
d^{2} r {\cal L} } , \nn && {\cal L} =
\psi_{\alpha}^{\dagger}[(\partial_{\tau} - \mu - 2 t)
\delta_{\alpha\beta} +
U_{\alpha\gamma}^{\dagger}\partial_{\tau}U_{\gamma\beta}]
\psi_{\beta} + t \Bigl(\partial_{\mathbf{r}}
\psi_{\beta}^{\dagger} - \psi_{\alpha}^{\dagger}
U_{\alpha\gamma}^{\dagger} (\partial_{\mathbf{r}} U_{\gamma\beta})
\Bigr) \Bigl(\partial_{\mathbf{r}}\psi_{\beta} -
(\partial_{\mathbf{r}}U^{\dagger})_{\beta\gamma}U_{\gamma\alpha}\psi_{\alpha}
\Bigr) - J_{H} S \sigma \psi_{\sigma}^{\dagger} \psi_{\sigma} \nn
&& + t \psi_{\alpha}^{\dagger}
\Bigl(\partial_{\mathbf{r}}U^{\dagger} +
[U^{\dagger}\partial_{\mathbf{r}}U]U^{\dagger}\Bigr)_{\alpha\gamma}
\Bigl(\partial_{\mathbf{r}}U +
U[(\partial_{\mathbf{r}}U^{\dagger})U]\Bigr)_{\gamma\beta}
\psi_{\beta} - t \partial_{\mathbf{r}}
(\psi_{\alpha}^{\dagger}\psi_{\alpha}) \nn && + i \lambda_{so}
\Bigl( \psi_{\gamma}^{\dagger} U_{\gamma\alpha}^{\dagger}
\sigma^{\mathbf{a}}_{\alpha\beta} U_{\beta \delta}
(\partial_{\mathbf{a}} \psi_{\delta}) - (\partial_{\mathbf{a}}
\psi_{\gamma}^{\dagger}) U_{\gamma\alpha}^{\dagger}
\sigma^{\mathbf{a}}_{\alpha\beta} U_{\beta \delta} \psi_{\delta} +
\psi_{\gamma}^{\dagger} U_{\gamma\alpha}^{\dagger}
\sigma^{\mathbf{a}}_{\alpha\beta} U_{\beta\xi}
U^{\dagger}_{\xi\chi} (\partial_{\mathbf{a}} U_{\chi \delta})
\psi_{\delta} - \psi_{\gamma}^{\dagger} (\partial_{\mathbf{a}}
U_{\gamma\alpha}^{\dagger}) U_{\alpha \xi} U^{\dagger}_{\xi\chi}
\sigma^{\mathbf{a}}_{\chi\beta} U_{\beta \delta} \psi_{\delta}
\Bigr) . \nn \eqa \end{widetext}

Introducing the Berry gauge field as the spin connection,
$\mathcal{A}_{\alpha\beta}^{\nu} = - i
[\{\partial_{\nu}U^{\dagger}\}U]_{\alpha\beta}$, we can rewrite
the above as follows \begin{widetext} \bqa && Z = \int D
\psi_{\alpha} D U_{\alpha\beta} D \mathcal{A}_{\alpha\beta}^{\nu}
\delta(U^{\dagger}_{\alpha\gamma}U_{\gamma\beta} -
\delta_{\alpha\beta}) \delta(\mathcal{A}_{\alpha\beta}^{\nu} + i
[\{\partial_{\nu}U^{\dagger}\}U]_{\alpha\beta})
\delta(\partial_{r}\mathcal{A}_{\alpha\beta}^{r}) e^{ - S_{B} -
\int_{0}^{\beta} d \tau \int d^{2} r {\cal L} } , \nn && {\cal L}
= \psi_{\alpha}^{\dagger}[(\partial_{\tau} - \mu - 2 t - 2
\lambda_{so}^{2}/t - J_{H} S \alpha) \delta_{\alpha\beta} - i
\mathcal{A}_{\alpha\beta}^{\tau}] \psi_{\beta} \nn && + t
\Bigl(\partial_{\mathbf{r}} \psi_{\beta}^{\dagger} + i
\psi_{\alpha}^{\dagger} [\mathcal{A}_{\alpha\beta}^{\mathbf{r}} +
(\lambda_{so}/t) U_{\alpha\gamma}^{\dagger}
\sigma^{\mathbf{r}}_{\gamma\delta} U_{\delta \beta}] \Bigr)
\Bigl(\partial_{\mathbf{r}}\psi_{\beta} - i [
\mathcal{A}_{\beta\gamma}^{\mathbf{r}} + (\lambda_{so}/t)
U_{\beta\alpha}^{\dagger} \sigma^{\mathbf{r}}_{\alpha\delta}
U_{\delta\gamma }] \psi_{\gamma} \Bigr) \nn && + t
\psi_{\alpha}^{\dagger}
\Bigl(\partial_{\mathbf{r}}U^{\dagger}_{\alpha\gamma} - i
[\mathcal{A}_{\alpha\delta}^{\mathbf{r}} + (\lambda_{so}/t)
U_{\alpha\beta}^{\dagger} \sigma^{\mathbf{r}}_{\beta\chi}
U_{\chi\delta }] U^{\dagger}_{\delta\gamma} \Bigr)
\Bigl(\partial_{\mathbf{r}}U_{\gamma\beta} + i U_{\gamma\xi}
[\mathcal{A}_{\xi\beta}^{\mathbf{r}} + (\lambda_{so}/t)
U_{\xi\gamma}^{\dagger} \sigma^{\mathbf{r}}_{\gamma\delta}
U_{\delta\beta }] \Bigr) \psi_{\beta} . \eqa \end{widetext}
Remember that this effective field theory is just the change of
variables in the microscopic model Eq. (A1).

Because this SU(2) gauge theory formulation is quite complicated,
we perform the U(1) approximation. One can understand this
procedure as the staggered-flux ansatz in the SU(2) slave-boson
theory, where the SU(2) gauge symmetry is reduced to the U(1)
symmetry \cite{Lee_Wen_Nagaosa}. Then, we find an effective U(1)
gauge theory \cite{Kim_Kim} \begin{widetext}\bqa && Z = \int D
\psi_{\sigma} D z_{\sigma} D a_{\mu} \delta(|z_{\sigma}|^{2} - 1)
\delta(\partial_{r}a_{r}) e^{ - S_{B} - \int_{0}^{\beta} d \tau
\int d^{2} r {\cal L} } , ~~~~~ {\cal L} = {\cal L}_{\psi} + {\cal
L}_{z} , \nn && {\cal L}_{\psi} = \psi_{\sigma}^{\dagger} (
\partial_{\tau} - \mu_{r} - J_{H} S \sigma - i \sigma a_{\tau} - i A_{\tau})
\psi_{\sigma} + t |[\partial_{\mathbf{r}} - i \sigma
a_{\mathbf{r}} - i A_{\mathbf{r}} - i \sigma (\lambda_{so}/t)
z_{\alpha}^{\dagger} \sigma^{\mathbf{r}}_{\alpha\beta} z_{\beta}]
\psi_{\sigma}|^{2} , \nn && {\cal L}_{z} = \rho_{s}
z_{\sigma}^{\dagger}(\partial_{\tau} - i a_{\tau})z_{\sigma} + t
\rho_{s} |[\partial_{\mathbf{r}} - i a_{\mathbf{r}} - i
(\lambda_{so}/t) z_{\alpha}^{\dagger}
\sigma^{\mathbf{r}}_{\alpha\beta} z_{\beta}]z_{\sigma}|^{2} , \eqa
\end{widetext} where an electromagnetic vector potential
$A_{\mathbf{r}}$ is introduced. It is interesting to see that the
internal gauge field $a_{\mathbf{r}}$ couples to the spin current
of the  electron field while the electromagnetic field
$A_{\mathbf{r}}$ does to the charge current. $\mu_{r} = \mu + 2 t
+ 2 \lambda_{so}^{2}/t$ is an effective chemical potential for
 itinerant electrons. The spinon part is reduced to the CP$^{1}$
representation of the ferromagnetic O(3) nonlinear $\sigma$ model,
where the time derivative term is added explicitly. This time
derivative term is expected to appear from quantum corrections,
i.e., the self-energy correction to the spinon dynamics. In the
antiferromagnetic case the second order time derivative term can
be applied to. The Rashba spin-orbit coupling gives rise to an
interaction term between the spin current and spin, quenching the
spin direction to the momentum or current direction.

Considering that the Berry phase term is associated with a
background potential, we are allowed to take into account the
saddle-point configuration of the gauge field for the Berry phase
term \bqa && a_{\mu} = - \frac{i}{2} [ z_{\sigma}^{\dagger}
(\partial_{\mu} z_{\sigma} ) - (\partial_{\mu}
z_{\sigma}^{\dagger}) z_{\sigma} ] = - \frac{\partial_{\mu}
\phi}{2} \cos \theta . \nonumber \eqa Then, the Berry phase term
can be written as follows \bqa && S_{B} = 2 i S \int_{0}^{\beta} d
\tau \int d^{2} r a_{\tau} , \eqa where $i S \int_{0}^{\beta} d
\tau \int d^{2} r \partial_{\tau} \phi = 0$ is used for the
skyrmion configuration.

\begin{widetext}

\section{Single particle Green's function in the single skyrmion
potential}

\subsection{Schrodinger equation with the single skyrmion
potential}

We start from the Schrodinger equation with the single skyrmion
potential \bqa (- i \partial_{t} - \mu_{r} - J_{H} S \sigma - i
\sigma a_{\tau}) \psi_{\sigma} - t (\partial_{\mathbf{r}} - i
\sigma a_{\mathbf{r}} - i A_{\mathbf{r}})^{2} \psi_{\sigma} = 0 ,
\eqa where the spin-orbit interaction is neglected. Introducing
the polar coordinate associated with the symmetry of the skyrmion
potential, \bqa && - t \Bigl(\boldsymbol{\hat{r}}
\partial_{r} + \boldsymbol{\hat{\phi}} \frac{\partial_{\phi}}{r} -
i \sigma a(r) \boldsymbol{\hat{\phi}} \Bigr) \cdot
\Bigl(\boldsymbol{\hat{r}}
\partial_{r} + \boldsymbol{\hat{\phi}} \frac{\partial_{\phi}}{r} - i \sigma
a(r) \boldsymbol{\hat{\phi}} \Bigr) = - t \Bigl(
\partial_{r}^{2} + \frac{\partial_{r}}{r} +
\frac{\partial_{\phi}^{2}}{r^{2}} - 2 i \sigma a(r)
\frac{\partial_{\phi}}{r} - a^{2}(r) \Bigr) , \eqa where the gauge
potential $a(r) \boldsymbol{\hat{\phi} } = a_{x}(x,y)
\boldsymbol{\hat{x}} + a_{y}(x,y) \boldsymbol{\hat{y}}$ in the
polar coordinate is given by \bqa && a(r) = \sqrt{ a_{x}^{2} +
a_{y}^{2}} = \frac{r}{\xi^{2} + r^{2}} \eqa for the single
skyrmion solution, we rewrite the above Schrodinger equation as
follows \bqa && ( - i \partial_{t} - \mu_{r} - J_{H} S \sigma )
\psi_{\sigma}(r, \phi, t) - t \Bigl\{
\partial_{r}^{2} + \frac{\partial_{r}}{r} + \frac{1}{r^{2}} \Bigl(
\partial_{\phi} + i \sigma \frac{ r^{2} }{\xi^{2} + r^{2}}
\Bigr)^{2} \Bigr\} \psi_{\sigma}(r, \phi, t) = 0 . \eqa

\subsection{Heun's confluent equation}

Inserting Eq. (\ref{Expansion}) with Eq. (\ref{Ansatz}) into Eq.
(B4), we obtain the eigen value problem for the radial part \bqa
&&
\partial_{r}^{2} F_{\sigma}^{n}(r) + \frac{\partial_{r}}{r}
F_{\sigma}^{n}(r) - \frac{1}{r^{2}} \Bigl( n + \sigma \frac{ r^{2}
}{\xi^{2} + r^{2}} \Bigr)^{2} F_{\sigma}^{n}(r) + t^{-1} (
E_{\sigma} + \mu_{r} + J_{H} S \sigma ) F_{\sigma}^{n}(r) = 0 .
\eqa Performing the change of variables $r^{2} = - \xi^{2} t$ and
introducing $F_{\sigma}^{n}(t) = t^{p} (t-1)^{q}
Y_{\sigma}^{n}(t)$ with constants $p$ and $q$, this equation can
be written as follows \bqa &&
\partial_{t}^{2} Y_{\sigma}^{n}(t) + \Bigl( \frac{2p+1}{t} +
\frac{2q}{t-1} \Bigr)
\partial_{t} Y_{\sigma}^{n}(t) + \Bigl(
\frac{p^{2} - n^{2}/4}{t^{2}} + \frac{(2 p + 1) q - n \sigma /2}{
t(t-1)} + \frac{q (q-1) - 1/4}{ (t-1)^{2}} - \frac{\xi^{2}}{4t}
\mathcal{E}_{\sigma} \Bigr) Y_{\sigma}^{n}(t) = 0 , \eqa where
$\mathcal{E}_{\sigma} \equiv t^{-1} (E_{\sigma} + \mu_{r} + J_{H}
S \sigma)$ is an effective energy level. Surprisingly, this
equation is known to be the Heun's confluent equation, and its
solution is well understood as follows \bqa Y_{\sigma}^{n}(t) =
\mbox{HeunC}\Bigl(0,n,\sqrt{2},\frac{1+n}{2}-\frac{\xi^{2}}{4 }
\mathcal{E}_{\sigma}, \frac{1- \sigma n}{2} + \frac{\xi^{2}}{4 }
\mathcal{E}_{\sigma}, t\Bigr) . \eqa

A general expression of the Heun's confluent equation \cite{Heun}
is given by \bqa && \frac{d^{2} Y(z)}{d z^{2}} - \frac{[-\alpha
z^{2} + (\alpha-\beta-\gamma-2)z+\beta+1]}{z(z-1)} \frac{d Y(z)}{d
z} - \frac{[(-\beta-\gamma-2)\alpha-2\delta]z + (\beta+1)z +
(-\gamma-1)\beta-\gamma-2\eta}{2z(z-1)} Y(z) = 0 , \nn \eqa
satisfying two boundary conditions such as \bqa && Y(z = 0) = 1,
\nn && \frac{d Y(z)}{d z}\Bigr|_{z = 0} =
\frac{(-\alpha+\gamma+1)\beta + \gamma-\alpha+2\eta}{2(\beta+1)} .
\eqa A general solution of this equation is known to be the Heun's
confluent function \bqa && Y(z) =
\mbox{HeunC}\Bigl(\alpha,\beta,\gamma,\delta,\eta;z \Bigr) , \eqa
where $p$ and $q$ are determined by \bqa && p^{2} =
\frac{n^{2}}{4} , ~~~~~ q(q-1) = \frac{1}{4} , \eqa and all other
coefficients are given by $p$ and $q$ in the following way \bqa &&
\alpha = 0 , ~~~~~ \beta = 2 p , ~~~~~ \gamma = 2 q - 1 , ~~~~~
\delta = p + \frac{1}{2} - \frac{\xi^{2}}{4} \mathcal{E}_{\sigma}
, ~~~~~ \eta = \frac{1-n\sigma}{2} + \frac{\xi^{2}}{4}
\mathcal{E}_{{\sigma}} . \eqa Two independent quantum numbers
appear to be an angular momentum $n$ and an energy eigen value
$\mathcal{E}_{\sigma}$, where $n$ is an integer while
$\mathcal{E}_{\sigma}$ turns out to be continuous.

In order to understand time reversal symmetry breaking in the
Heun's confluent function, it is valuable to find its asymptotic
form in the $r \rightarrow \infty$ limit. The corresponding
Schrodinger equation is given by \bqa && \frac{2p + 2q +1}{t}
\partial_{t} Y_{\sigma}^{n}(t) - \frac{\xi^{2}}{4t}
\mathcal{E}_{\sigma} Y_{\sigma}^{n}(t) \approx 0 , \eqa where
dominant terms are selected by the $t \rightarrow - \infty$ limit.
It is straightforward to solve this differential equation.
Introducing $p = n/2$ and $q = (1+\sqrt{2})/2$ into the solution,
we obtain \bqa && Y_{\sigma}^{n}(r \rightarrow \infty) \propto
\exp\Bigl( - \frac{\mathcal{E}_{\sigma}}{4(n+2+\sqrt{2})} r^{2}
\Bigr) . \eqa

\subsection{A Green's function in a skyrmion background}

Inserting Eq. (\ref{Expansion}) with Eq. (\ref{Ansatz}) into the
Green's function, we obtain \bqa &&
G_{\sigma\sigma}(r,r',\phi-\phi',t-t')  = - |\mathcal{C}|^{2} \int
d \mathcal{E}_{\sigma} e^{- i \mathcal{E}_{\sigma} (t-t')}
\sum_{n} e^{in(\phi-\phi')} F_{\sigma}^{n}(r;\mathcal{E}_{\sigma})
F_{\sigma}^{n}(r';\mathcal{E}_{\sigma}) \Bigl\langle
a_{\sigma}(n,\mathcal{E}_{\sigma})
a_{\sigma}^{\dagger}(n,\mathcal{E}_{\sigma}) \Bigr\rangle . \eqa
We note that the radial coordinate cannot be $r-r'$ due to
translational symmetry breaking.

Performing the Fourier transformation for time, we obtain \bqa &&
G_{\sigma\sigma}(r,r',\phi-\phi',\omega+i\delta) =
|\mathcal{C}|^{2} \int d \mathcal{E}_{\sigma} \sum_{n}
e^{in(\phi-\phi')} F_{\sigma}^{n}(r;\mathcal{E}_{\sigma})
F_{\sigma}^{n}(r';\mathcal{E}_{\sigma}) \Bigl(
\frac{f(\mathcal{E}_{\sigma})}{\omega - \mathcal{E}_{\sigma} -
i\delta} + \frac{1-f(\mathcal{E}_{\sigma})}{\omega -
\mathcal{E}_{\sigma} + i\delta} \Bigr)  \eqa in the real
frequency, where $f(\mathcal{E}_{\sigma})$ is the Fermi-Dirac
distribution function, while given by \bqa &&
G_{\sigma\sigma}(r,r',\phi-\phi',i\omega) = |\mathcal{C}|^{2} \int
d \mathcal{E}_{\sigma} \sum_{n} e^{in(\phi-\phi')} \frac{
F_{\sigma}^{n}(r;\mathcal{E}_{\sigma})
F_{\sigma}^{n}(r';\mathcal{E}_{\sigma}) }{i\omega -
\mathcal{E}_{\sigma} }  \eqa in the Matsubara frequency.

We introduce the self-energy correction, resulting from both
elastic impurity scattering and inelastic interaction effects with
gauge fluctuations \bqa &&
\Sigma_{\sigma}(i\omega,\mathcal{E}_{\sigma}) =
\Sigma_{\sigma}^{imp}(i\omega,\mathcal{E}_{\sigma}) +
\Sigma_{\sigma}^{ele}(i\omega,\mathcal{E}_{\sigma})
. \eqa Then, the most general expression for the single particle
Green's function is given by \bqa &&
G_{\sigma\sigma}(r,r',\phi-\phi',i\omega) = |\mathcal{C}|^{2} \int
d \mathcal{E}_{\sigma} \sum_{n} e^{in(\phi-\phi')}
\frac{F_{\sigma}^{n}(r;\mathcal{E}_{\sigma})
F_{\sigma}^{n}(r';\mathcal{E}_{\sigma})}{i\omega -
\Sigma_{\sigma}(i\omega,\mathcal{E}_{\sigma}) -
\mathcal{E}_{\sigma} } . \eqa

\section{Polarization functions for longitudinal conductivity}

\subsection{Formal expressions}

For convenience, we decompose $\Pi_{xx}^{ss}$ into four contributions;
\bqa
&& \Pi_{xx}^{ss}(r,r',\phi-\phi',i\Omega) =
\sum_{K=A,B,C,D}\Pi_{xx}^{ss(K)}(r,r',\phi-\phi',i\Omega).
\eqa

The first part is the conventional particle-hole channel in the
presence of the single skyrmion potential, \bqa &&
\Pi_{xx}^{ss(A)}(r,r',\phi-\phi',i\Omega) = - t^{2}
|\mathcal{C}|^{4} \cos \phi \cos \phi' \sum_{n} \sum_{n'}
e^{i(n-n')(\phi-\phi')} \int d \mathcal{E}_{\sigma} \int d
\mathcal{E}'_{\sigma} \nn && \Bigl\{ [\partial_{r}
F_{\sigma}^{n}(r;\mathcal{E}_{\sigma})]
F_{\sigma}^{n}(r';\mathcal{E}_{\sigma}) [\partial_{r'}
F_{\sigma}^{n'}(r';\mathcal{E}'_{\sigma})]
F_{\sigma}^{n'}(r;\mathcal{E}'_{\sigma}) - [\partial_{r}
F_{\sigma}^{n}(r;\mathcal{E}_{\sigma})] [\partial_{r'}
F_{\sigma}^{n}(r';\mathcal{E}_{\sigma})]
F_{\sigma}^{n'}(r';\mathcal{E}'_{\sigma})
F_{\sigma}^{n'}(r;\mathcal{E}'_{\sigma}) \nn && -
F_{\sigma}^{n}(r;\mathcal{E}_{\sigma})
F_{\sigma}^{n}(r';\mathcal{E}_{\sigma}) [\partial_{r'}
F_{\sigma}^{n'}(r';\mathcal{E}'_{\sigma})] [\partial_{r}
F_{\sigma}^{n'}(r;\mathcal{E}'_{\sigma}) ] +
F_{\sigma}^{n}(r;\mathcal{E}_{\sigma}) [\partial_{r'}
F_{\sigma}^{n}(r';\mathcal{E}_{\sigma})]
F_{\sigma}^{n'}(r';\mathcal{E}'_{\sigma}) [\partial_{r}
F_{\sigma}^{n'}(r;\mathcal{E}'_{\sigma})] \Bigr\} \nn &&
\int_{-\infty}^{\infty} d \nu_{\sigma} \int_{-\infty}^{\infty} d
\nu_{\sigma}' \frac{f(\nu_{\sigma}) - f(\nu_{\sigma}')}{i\Omega -
(\nu_{\sigma} -\nu_{\sigma}')} {\cal
A}_{\sigma}(\nu_{\sigma},\mathcal{E}_{\sigma}) {\cal
A}_{\sigma}(\nu_{\sigma}',\mathcal{E}_{\sigma}') , \eqa where the
spectral function for itinerant electrons is given by Eq.
(\ref{Spectral_FT}). In the ideal non-interacting limit
$\Re\Sigma_{\sigma}(\nu_{\sigma}),
\Im\Sigma_{\sigma}(\nu_{\sigma}) \rightarrow 0$ we obtain \bqa &&
\int_{-\infty}^{\infty} d \nu_{\sigma} \int_{-\infty}^{\infty} d
\nu_{\sigma}' \frac{f(\nu_{\sigma}) - f(\nu_{\sigma}')}{i\Omega -
(\nu_{\sigma} -\nu_{\sigma}')} {\cal
A}_{\sigma}(\nu_{\sigma},\mathcal{E}_{\sigma}) {\cal
A}_{\sigma}(\nu_{\sigma}',\mathcal{E}_{\sigma}') \longrightarrow
\frac{f(\mathcal{E}_{\sigma}) - f(\mathcal{E}'_{\sigma})}{i\Omega
- (\mathcal{E}_{\sigma} - \mathcal{E}'_{\sigma})} , \eqa nothing
but the particle-hole polarization function in the Fermi gas.

Other pieces are given by
\bqa && \Pi_{xx}^{ss(B)}(r,r',\phi-\phi',i\Omega)
= t^{2} |\mathcal{C}|^{4} \frac{\sin \phi}{r} \cos \phi' \sum_{n}
\sum_{n'} i (n-n') e^{i(n-n')(\phi-\phi')} \int d
\mathcal{E}_{\sigma} \int d \mathcal{E}'_{\sigma} \nn && \Bigl\{
F_{\sigma}^{n}(r;\mathcal{E}_{\sigma})
F_{\sigma}^{n}(r';\mathcal{E}_{\sigma}) [\partial_{r'}
F_{\sigma}^{n'}(r';\mathcal{E}'_{\sigma})]
F_{\sigma}^{n'}(r;\mathcal{E}'_{\sigma}) -
F_{\sigma}^{n}(r;\mathcal{E}_{\sigma}) [\partial_{r'}
F_{\sigma}^{n}(r';\mathcal{E}_{\sigma})]
F_{\sigma}^{n'}(r';\mathcal{E}'_{\sigma})
F_{\sigma}^{n'}(r;\mathcal{E}'_{\sigma}) \nn && -
F_{\sigma}^{n}(r;\mathcal{E}_{\sigma})
F_{\sigma}^{n}(r';\mathcal{E}_{\sigma})
F_{\sigma}^{n'}(r';\mathcal{E}'_{\sigma}) [\partial_{r}
F_{\sigma}^{n'}(r;\mathcal{E}'_{\sigma}) ] +
F_{\sigma}^{n}(r;\mathcal{E}_{\sigma})
F_{\sigma}^{n}(r';\mathcal{E}_{\sigma})
F_{\sigma}^{n'}(r';\mathcal{E}'_{\sigma}) [\partial_{r}
F_{\sigma}^{n'}(r;\mathcal{E}'_{\sigma})] \Bigr\} \nn &&
\int_{-\infty}^{\infty} d \nu_{\sigma} \int_{-\infty}^{\infty} d
\nu_{\sigma}' \frac{f(\nu_{\sigma}) - f(\nu_{\sigma}')}{i\Omega -
(\nu_{\sigma} -\nu_{\sigma}')} {\cal
A}_{\sigma}(\nu_{\sigma},\mathcal{E}_{\sigma}) {\cal
A}_{\sigma}(\nu_{\sigma}',\mathcal{E}_{\sigma}') , \eqa
\bqa && \Pi_{xx}^{ss(C)}(r,r',\phi-\phi',i\Omega) = - t^{2}
|\mathcal{C}|^{4} \cos \phi \frac{\sin \phi'}{r'} \sum_{n}
\sum_{n'} i (n-n') e^{i(n-n')(\phi-\phi')} \int d
\mathcal{E}_{\sigma} \int d \mathcal{E}'_{\sigma} \nn && \Bigl\{
[\partial_{r} F_{\sigma}^{n}(r;\mathcal{E}_{\sigma})]
F_{\sigma}^{n}(r';\mathcal{E}_{\sigma})
F_{\sigma}^{n'}(r';\mathcal{E}'_{\sigma})
F_{\sigma}^{n'}(r;\mathcal{E}'_{\sigma}) - [\partial_{r}
F_{\sigma}^{n}(r;\mathcal{E}_{\sigma})]
F_{\sigma}^{n}(r';\mathcal{E}_{\sigma})
F_{\sigma}^{n'}(r';\mathcal{E}'_{\sigma})
F_{\sigma}^{n'}(r;\mathcal{E}'_{\sigma}) \nn && -
F_{\sigma}^{n}(r;\mathcal{E}_{\sigma})
F_{\sigma}^{n}(r';\mathcal{E}_{\sigma}) [\partial_{r'}
F_{\sigma}^{n'}(r';\mathcal{E}'_{\sigma}) ]
F_{\sigma}^{n'}(r;\mathcal{E}'_{\sigma}) +
F_{\sigma}^{n}(r;\mathcal{E}_{\sigma}) [\partial_{r'}
F_{\sigma}^{n}(r';\mathcal{E}_{\sigma})]
F_{\sigma}^{n'}(r';\mathcal{E}'_{\sigma})
F_{\sigma}^{n'}(r;\mathcal{E}'_{\sigma}) \Bigr\} \nn &&
\int_{-\infty}^{\infty} d \nu_{\sigma} \int_{-\infty}^{\infty} d
\nu_{\sigma}' \frac{f(\nu_{\sigma}) - f(\nu_{\sigma}')}{i\Omega -
(\nu_{\sigma} -\nu_{\sigma}')} {\cal
A}_{\sigma}(\nu_{\sigma},\mathcal{E}_{\sigma}) {\cal
A}_{\sigma}(\nu_{\sigma}',\mathcal{E}_{\sigma}') , \eqa and \bqa
&& \Pi_{xx}^{ss(D)}(r,r',\phi-\phi',i\Omega) = - t^{2}
|\mathcal{C}|^{4} \frac{\sin \phi}{r} \frac{\sin \phi'}{r'}
\sum_{n} \sum_{n'} (n-n')^{2} e^{i(n-n')(\phi-\phi')} \int d
\mathcal{E}_{\sigma} \int d \mathcal{E}'_{\sigma} \nn && \Bigl\{
F_{\sigma}^{n}(r;\mathcal{E}_{\sigma})
F_{\sigma}^{n}(r';\mathcal{E}_{\sigma})
F_{\sigma}^{n'}(r';\mathcal{E}'_{\sigma})
F_{\sigma}^{n'}(r;\mathcal{E}'_{\sigma}) -
F_{\sigma}^{n}(r;\mathcal{E}_{\sigma})
F_{\sigma}^{n}(r';\mathcal{E}_{\sigma})
F_{\sigma}^{n'}(r';\mathcal{E}'_{\sigma})
F_{\sigma}^{n'}(r;\mathcal{E}'_{\sigma}) \nn && -
F_{\sigma}^{n}(r;\mathcal{E}_{\sigma})
F_{\sigma}^{n}(r';\mathcal{E}_{\sigma})
F_{\sigma}^{n'}(r';\mathcal{E}'_{\sigma})
F_{\sigma}^{n'}(r;\mathcal{E}'_{\sigma}) +
F_{\sigma}^{n}(r;\mathcal{E}_{\sigma})
F_{\sigma}^{n}(r';\mathcal{E}_{\sigma})
F_{\sigma}^{n'}(r';\mathcal{E}'_{\sigma})
F_{\sigma}^{n'}(r;\mathcal{E}'_{\sigma}) \Bigr\} \nn &&
\int_{-\infty}^{\infty} d \nu_{\sigma} \int_{-\infty}^{\infty} d
\nu_{\sigma}' \frac{f(\nu_{\sigma}) - f(\nu_{\sigma}')}{i\Omega -
(\nu_{\sigma} -\nu_{\sigma}')} {\cal
A}_{\sigma}(\nu_{\sigma},\mathcal{E}_{\sigma}) {\cal
A}_{\sigma}(\nu_{\sigma}',\mathcal{E}_{\sigma}') . \eqa

Substituting the above expressions into Eq.
(\ref{Longitudinal_Conductivity}), it is found that the
contributions from $\Pi_{xx}^{ss(D)}(r,r',\phi-\phi',i\Omega)$ and
$\Pi_{xx}^{ss(B)}(r,r',\phi-\phi',i\Omega)+\Pi_{xx}^{ss(C)}(r,r',\phi-\phi',i\Omega)$
all vanish. Therefore, we obtain Eq. (\ref{Sigma_XX}).

\subsection{Asymptotic form}

The conductivity is simplified as follows in the non-interacting
limit without the spin-orbit interaction \bqa &&
\sigma_{sp}^{\psi} = 2 \pi^{2} t^{2} |\mathcal{C}|^{4} \int d
\mathcal{E}_{\sigma} \Bigl( - \frac{\partial f(\Omega)}{\partial
\Omega} \Bigr)_{\Omega = \mathcal{E}_{\sigma}} \sum_{n}
[\mathcal{R}_{\sigma}^n(\mathcal{E}_{\sigma};\mathcal{E}_{\sigma})]^{2}
. \eqa

%
%

Resorting to the asymptotic expression, we see \bqa &&
\partial_{r} F_{\sigma}^{n}(r\rightarrow
\infty;\mathcal{E}_{\sigma}) = \Bigl( (1+\sqrt{2})
\frac{r}{\xi^{2} + r^{2}} + \frac{n}{r} -
\frac{\mathcal{E}_{\sigma}}{2(n+2+\sqrt{2})} r \Bigr)
F_{\sigma}^{n}(r\rightarrow \infty;\mathcal{E}_{\sigma}) . \eqa
Then, we obtain \bqa &&
\mathcal{R}_{\sigma}^n(\mathcal{E}_{\sigma};\mathcal{E}_{\sigma})
= - \int_{0}^{\infty} dr r \Bigl( \frac{1}{r} +
\frac{\mathcal{E}_{\sigma}}{2(n+2+\sqrt{2})(n+3+\sqrt{2})} r
\Bigr) F_{\sigma}^{n}(r;\mathcal{E}_{\sigma})
F_{\sigma}^{n+1}(r;\mathcal{E}_{\sigma}) \nn && = -
(\mathcal{C}'_{1})^{2} \int_{0}^{\infty} dr r \Bigl( \frac{1}{r} +
\frac{\mathcal{E}_{\sigma}}{2(n+2+\sqrt{2})(n+3+\sqrt{2})} r
\Bigr) (\xi^{2} + r^{2})^{1+\sqrt{2}} r^{2n+1} \exp\Bigl( -
\mathcal{E}_{\sigma} \frac{2n + 5 + 2\sqrt{2}}{4(n + 2 +
\sqrt{2})(n + 3 + \sqrt{2})} r^{2} \Bigr) . \nn \eqa

Inserting this expression into Eq. (C7), we reach the following
expression in the $T \rightarrow 0$ limit \bqa
\sigma_{sp}^{\psi}(T \rightarrow 0) &=& 2 \pi^{2} t^{2}
|\mathcal{C}|^{4} (\mathcal{C}'_{1})^{4} \int d
\mathcal{E}_{\sigma} \delta(\mathcal{E}_{\sigma}) \sum_{n} \Bigl\{
\int_{0}^{L} dr r \Bigl( \frac{1}{r} +
\frac{\mathcal{E}_{\sigma}}{2(n+2+\sqrt{2})(n+3+\sqrt{2})} r
\Bigr) \nn &\times& (\xi^{2} + r^{2})^{1+\sqrt{2}} r^{2n+1}
\exp\Bigl( - \mathcal{E}_{\sigma} \frac{2n + 5 + 2\sqrt{2}}{4(n +
2 + \sqrt{2})(n + 3 + \sqrt{2})} r^{2} \Bigr) \Bigr\}^{2} \nn &=&
2 \pi^{2} t^{2} |\mathcal{C}|^{4} (\mathcal{C}'_{1})^{4} \sum_{n}
\xi^{4+2\sqrt{2}+2n} \Bigl\{ \int_{0}^{L/\xi} d x x^{2n+1}
(1+x^{2})^{1+\sqrt{2}} \Bigr\} \nn &=& 2 \pi^{2} t^{2}
|\mathcal{C}|^{4} (\mathcal{C}'_{1})^{4} \sum_{n}
\xi^{4+2\sqrt{2}+2n} \Bigl( \frac{L}{\xi} \Bigr)^{2n+2} \Bigl\{
\frac{{}_2F_1[1 + n, -\sqrt{2}, 2 + n, -(L/\xi)^2]}{2(n+1)} \nn
&+& \Bigl( \frac{L}{\xi} \Bigr)^{2} \frac{{}_2F_1[2 + n,
-\sqrt{2}, 3 + n, -(L/\xi)^2]}{2(n+2)} \Bigr\} , \eqa where
$L/\xi$ is the ratio between the system size and the skyrmion core
size and ${}_2F_1$ is the hypergeometric function.

\section{Polarization functions for Hall conductivity}

\subsection{Formal expressions}

It is convenient to decompose $\Pi_{xy}^{ss}$ into four contributions;
\bqa &&
\Pi_{xy}^{ss}(r,r',\phi-\phi',i\Omega) =
\sum_{K=A,B,C,D}\Pi_{xy}^{ss(K)}(r,r',\phi-\phi',i\Omega).
\eqa
The four polarization functions are given by
\bqa &&
\Pi_{xy}^{ss(A)}(r,r',\phi-\phi',i\Omega) = - t^{2}
|\mathcal{C}|^{4} \cos \phi \sin \phi' \sum_{n} \sum_{n'}
e^{i(n-n')(\phi-\phi')} \int d \mathcal{E}_{\sigma} \int d
\mathcal{E}'_{\sigma} \nn && \Bigl\{ [\partial_{r}
F_{\sigma}^{n}(r;\mathcal{E}_{\sigma})]
F_{\sigma}^{n}(r';\mathcal{E}_{\sigma}) [\partial_{r'}
F_{\sigma}^{n'}(r';\mathcal{E}'_{\sigma})]
F_{\sigma}^{n'}(r;\mathcal{E}'_{\sigma}) - [\partial_{r}
F_{\sigma}^{n}(r;\mathcal{E}_{\sigma})] [\partial_{r'}
F_{\sigma}^{n}(r';\mathcal{E}_{\sigma})]
F_{\sigma}^{n'}(r';\mathcal{E}'_{\sigma})
F_{\sigma}^{n'}(r;\mathcal{E}'_{\sigma}) \nn && -
F_{\sigma}^{n}(r;\mathcal{E}_{\sigma})
F_{\sigma}^{n}(r';\mathcal{E}_{\sigma}) [\partial_{r'}
F_{\sigma}^{n'}(r';\mathcal{E}'_{\sigma})] [\partial_{r}
F_{\sigma}^{n'}(r;\mathcal{E}'_{\sigma}) ] +
F_{\sigma}^{n}(r;\mathcal{E}_{\sigma}) [\partial_{r'}
F_{\sigma}^{n}(r';\mathcal{E}_{\sigma})]
F_{\sigma}^{n'}(r';\mathcal{E}'_{\sigma}) [\partial_{r}
F_{\sigma}^{n'}(r;\mathcal{E}'_{\sigma})] \Bigr\} \nn &&
\int_{-\infty}^{\infty} d \nu_{\sigma} \int_{-\infty}^{\infty} d
\nu_{\sigma}' \frac{f(\nu_{\sigma}) - f(\nu_{\sigma}')}{i\Omega -
(\nu_{\sigma} -\nu_{\sigma}')} {\cal
A}_{\sigma}(\nu_{\sigma},\mathcal{E}_{\sigma}) {\cal
A}_{\sigma}(\nu_{\sigma}',\mathcal{E}_{\sigma}') , \eqa
\bqa && \Pi_{xy}^{ss(B)}(r,r',\phi-\phi',i\Omega) = t^{2}
|\mathcal{C}|^{4} \frac{\sin \phi}{r} \sin \phi' \sum_{n}
\sum_{n'} i (n-n') e^{i(n-n')(\phi-\phi')} \int d
\mathcal{E}_{\sigma} \int d \mathcal{E}'_{\sigma} \nn && \Bigl\{
F_{\sigma}^{n}(r;\mathcal{E}_{\sigma})
F_{\sigma}^{n}(r';\mathcal{E}_{\sigma}) [\partial_{r'}
F_{\sigma}^{n'}(r';\mathcal{E}'_{\sigma})]
F_{\sigma}^{n'}(r;\mathcal{E}'_{\sigma}) -
F_{\sigma}^{n}(r;\mathcal{E}_{\sigma}) [\partial_{r'}
F_{\sigma}^{n}(r';\mathcal{E}_{\sigma})]
F_{\sigma}^{n'}(r';\mathcal{E}'_{\sigma})
F_{\sigma}^{n'}(r;\mathcal{E}'_{\sigma}) \nn && -
F_{\sigma}^{n}(r;\mathcal{E}_{\sigma})
F_{\sigma}^{n}(r';\mathcal{E}_{\sigma})
F_{\sigma}^{n'}(r';\mathcal{E}'_{\sigma}) [\partial_{r}
F_{\sigma}^{n'}(r;\mathcal{E}'_{\sigma}) ] +
F_{\sigma}^{n}(r;\mathcal{E}_{\sigma})
F_{\sigma}^{n}(r';\mathcal{E}_{\sigma})
F_{\sigma}^{n'}(r';\mathcal{E}'_{\sigma}) [\partial_{r}
F_{\sigma}^{n'}(r;\mathcal{E}'_{\sigma})] \Bigr\} \nn &&
\int_{-\infty}^{\infty} d \nu_{\sigma} \int_{-\infty}^{\infty} d
\nu_{\sigma}' \frac{f(\nu_{\sigma}) - f(\nu_{\sigma}')}{i\Omega -
(\nu_{\sigma} -\nu_{\sigma}')} {\cal
A}_{\sigma}(\nu_{\sigma},\mathcal{E}_{\sigma}) {\cal
A}_{\sigma}(\nu_{\sigma}',\mathcal{E}_{\sigma}') , \eqa
\bqa && \Pi_{xy}^{ss(C)}(r,r',\phi-\phi',i\Omega) = t^{2}
|\mathcal{C}|^{4} \cos \phi \frac{\cos \phi'}{r'} \sum_{n}
\sum_{n'} i(n-n') e^{i(n-n')(\phi-\phi')} \int d
\mathcal{E}_{\sigma} \int d \mathcal{E}'_{\sigma} \nn && \Bigl\{
[\partial_{r} F_{\sigma}^{n}(r;\mathcal{E}_{\sigma})]
F_{\sigma}^{n}(r';\mathcal{E}_{\sigma})
F_{\sigma}^{n'}(r';\mathcal{E}'_{\sigma})
F_{\sigma}^{n'}(r;\mathcal{E}'_{\sigma}) - [\partial_{r}
F_{\sigma}^{n}(r;\mathcal{E}_{\sigma})]
F_{\sigma}^{n}(r';\mathcal{E}_{\sigma})
F_{\sigma}^{n'}(r';\mathcal{E}'_{\sigma})
F_{\sigma}^{n'}(r;\mathcal{E}'_{\sigma}) \nn && -
F_{\sigma}^{n}(r;\mathcal{E}_{\sigma})
F_{\sigma}^{n}(r';\mathcal{E}_{\sigma}) [\partial_{r'}
F_{\sigma}^{n'}(r';\mathcal{E}'_{\sigma}) ]
F_{\sigma}^{n'}(r;\mathcal{E}'_{\sigma}) +
F_{\sigma}^{n}(r;\mathcal{E}_{\sigma}) [\partial_{r'}
F_{\sigma}^{n}(r';\mathcal{E}_{\sigma})]
F_{\sigma}^{n'}(r';\mathcal{E}'_{\sigma})
F_{\sigma}^{n'}(r;\mathcal{E}'_{\sigma}) \Bigr\} \nn &&
\int_{-\infty}^{\infty} d \nu_{\sigma} \int_{-\infty}^{\infty} d
\nu_{\sigma}' \frac{f(\nu_{\sigma}) - f(\nu_{\sigma}')}{i\Omega -
(\nu_{\sigma} -\nu_{\sigma}')} {\cal
A}_{\sigma}(\nu_{\sigma},\mathcal{E}_{\sigma}) {\cal
A}_{\sigma}(\nu_{\sigma}',\mathcal{E}_{\sigma}') , \eqa and \bqa
&& \Pi_{xy}^{ss(D)}(r,r',\phi-\phi',i\Omega) = t^{2}
|\mathcal{C}|^{4} \frac{\sin \phi}{r} \frac{\cos \phi'}{r'}
\sum_{n} \sum_{n'} (n-n')^{2} e^{i(n-n')(\phi-\phi')} \int d
\mathcal{E}_{\sigma} \int d \mathcal{E}'_{\sigma} \nn && \Bigl\{
F_{\sigma}^{n}(r;\mathcal{E}_{\sigma})
F_{\sigma}^{n}(r';\mathcal{E}_{\sigma})
F_{\sigma}^{n'}(r';\mathcal{E}'_{\sigma})
F_{\sigma}^{n'}(r;\mathcal{E}'_{\sigma}) -
F_{\sigma}^{n}(r;\mathcal{E}_{\sigma})
F_{\sigma}^{n}(r';\mathcal{E}_{\sigma})
F_{\sigma}^{n'}(r';\mathcal{E}'_{\sigma})
F_{\sigma}^{n'}(r;\mathcal{E}'_{\sigma}) \nn && -
F_{\sigma}^{n}(r;\mathcal{E}_{\sigma})
F_{\sigma}^{n}(r';\mathcal{E}_{\sigma})
F_{\sigma}^{n'}(r';\mathcal{E}'_{\sigma})
F_{\sigma}^{n'}(r;\mathcal{E}'_{\sigma}) +
F_{\sigma}^{n}(r;\mathcal{E}_{\sigma})
F_{\sigma}^{n}(r';\mathcal{E}_{\sigma})
F_{\sigma}^{n'}(r';\mathcal{E}'_{\sigma})
F_{\sigma}^{n'}(r;\mathcal{E}'_{\sigma}) \Bigr\} \nn &&
\int_{-\infty}^{\infty} d \nu_{\sigma} \int_{-\infty}^{\infty} d
\nu_{\sigma}' \frac{f(\nu_{\sigma}) - f(\nu_{\sigma}')}{i\Omega -
(\nu_{\sigma} -\nu_{\sigma}')} {\cal
A}_{\sigma}(\nu_{\sigma},\mathcal{E}_{\sigma}) {\cal
A}_{\sigma}(\nu_{\sigma}',\mathcal{E}_{\sigma}') . \eqa
It is found that $\Pi_{xy}^{ss(A)}(r,r',\phi-\phi',i\Omega)$ and
$\Pi_{xy}^{ss(D)}(r,r',\phi-\phi',i\Omega)$ do not contribute to
$\sigma_{sH}^{\psi}$, while
$\Pi_{xy}^{ss(B)}(r,r',\phi-\phi',i\Omega)$ and
$\Pi_{xy}^{ss(C)}(r,r',\phi-\phi',i\Omega)$ give finite
contributions.

\subsection{Asymptotic forms}

In the non-interacting limit the Hall conductivity is given by
\bqa && \sigma_{sH}^{\psi} = 2 \pi^{2} t^{2} |\mathcal{C}|^{4}
\int d \mathcal{E}_{\sigma} \int d
\mathcal{E}'_{\sigma}\frac{f(\mathcal{E}_{\sigma}) -
f(\mathcal{E}'_{\sigma})}{ (\mathcal{E}_{\sigma} -
\mathcal{E}'_{\sigma}) } \sum_{n} \Bigl\{
\mathcal{O}_{\sigma}^n(\mathcal{E}_{\sigma};\mathcal{E}_{\sigma}')
\mathcal{R}_{\sigma}^n(\mathcal{E}_{\sigma};\mathcal{E}_{\sigma}')
-
\mathcal{P}_{\sigma}^n(\mathcal{E}_{\sigma};\mathcal{E}_{\sigma}')
\mathcal{Q}_{\sigma}^n(\mathcal{E}_{\sigma};\mathcal{E}_{\sigma}')
\Bigr\} , \eqa where the spin-orbit interaction is not introduced.

Inserting the asymptotic expression of the radial wave function as
performed in appendix C2, we obtain \bqa
\mathcal{O}_{\sigma}^n(\mathcal{E}_{\sigma};\mathcal{E}_{\sigma}')
&=& (\mathcal{C}'_{1})^{2} \int_{0}^{\infty} dr (\xi^{2} +
r^{2})^{1+\sqrt{2}} r^{2n+1} \exp\Bigl( - \Bigl\{
\frac{\mathcal{E}_{\sigma}}{4(n + 2 + \sqrt{2})} +
\frac{\mathcal{E}'_{\sigma}}{4(n + 3 + \sqrt{2})} \Bigr\} r^{2}
\Bigr)  , \nn
\mathcal{P}_{\sigma}^n(\mathcal{E}_{\sigma};\mathcal{E}_{\sigma}')
&=& (\mathcal{C}'_{1})^{2} \int_{0}^{\infty} dr r (\xi^{2} +
r^{2})^{1+\sqrt{2}} r^{2n+1} \exp\Bigl( - \Bigl\{
\frac{\mathcal{E}_{\sigma}}{4(n + 2 + \sqrt{2})} +
\frac{\mathcal{E}'_{\sigma}}{4(n + 3 + \sqrt{2})} \Bigr\} r^{2}
\Bigr) , \nn
\mathcal{Q}_{\sigma}^n(\mathcal{E}_{\sigma};\mathcal{E}_{\sigma}')
&=& - (\mathcal{C}'_{1})^{2} \int_{0}^{\infty} dr \Bigl(
\frac{1}{r} + \frac{\mathcal{E}_{\sigma}}{2(n+2+\sqrt{2})} r -
\frac{\mathcal{E}'_{\sigma}}{2(n+3+\sqrt{2})} r \Bigr) \nn
&\times& (\xi^{2} + r^{2})^{1+\sqrt{2}} r^{2n+1} \exp\Bigl( -
\Bigl\{ \frac{\mathcal{E}_{\sigma}}{4(n + 2 + \sqrt{2})} +
\frac{\mathcal{E}'_{\sigma}}{4(n + 3 + \sqrt{2})} \Bigr\} r^{2}
\Bigr)   , \nn
\mathcal{R}_{\sigma}^n(\mathcal{E}_{\sigma};\mathcal{E}_{\sigma}')
&=& - (\mathcal{C}'_{1})^{2} \int_{0}^{\infty} dr r \Bigl(
\frac{1}{r} + \frac{\mathcal{E}_{\sigma}}{2(n+2+\sqrt{2})} r -
\frac{\mathcal{E}'_{\sigma}}{2(n+3+\sqrt{2})} r \Bigr) \nn
&\times& (\xi^{2} + r^{2})^{1+\sqrt{2}} r^{2n+1} \exp\Bigl( -
\Bigl\{ \frac{\mathcal{E}_{\sigma}}{4(n + 2 + \sqrt{2})} +
\frac{\mathcal{E}'_{\sigma}}{4(n + 3 + \sqrt{2})} \Bigr\} r^{2}
\Bigr) .  \eqa It is difficult to perform further simplification
analytically for the general case of $n$. However, it is clear
that this expression does not vanish due to the factor of $r$ in
the $r$ integration.

\end{widetext}


\begin{thebibliography}{9}
\bibitem{Conducting_Polymer} A. J. Heeger, S. Kivelson,
J. R. Schrieffer, and W. -P. Su, Rev. Mod. Phys. {\bf 60}, 781
(1988).
\bibitem{Vortex_Dynamics_Review} G. Blatter, M. V. Feigel'man,
V. B. Geshkenbein, A. I. Larkin, and V. M. Vinokur, Rev. Mod.
Phys. {\bf 66}, 1125 (1994).
\bibitem{Soliton_High_Energy} G. 't Hooft and F. Bruckmann,
arXiv:hep-th/0010225 (unpublished).
\bibitem{Skyrmion_Original} T. H. R. Skyrme, Nuclear Physics {\bf 31},
556 (1962).
\bibitem{Skyrmion_IQHE} S. L. Sondhi, A. Karlhede, S. A. Kivelson,
and E. H. Rezayi, Phys. Rev. B {\bf 47}, 16419 (1993).
\bibitem{Skyrmion_Magnon_IQHE} D. Fukuoka, K. Oto, K. Muro, Y. Hirayama, and N.
Kumada, Phys. Rev. Lett. {\bf 105}, 126802 (2010).
\bibitem{Skyrmion_Condensation} U. Al Khawaja and H. Stoof, Nature
{\bf 411}, 918 (2001).
\bibitem{MnSi_Skyrmion_Crystal} S. Muhlbauer, B. Binz, F. Jonietz,
C. Pfleiderer, A. Rosch, A. Neubauer, R. Georgii, and P. Boni, Science {\bf 323}, 915
(2009).
\bibitem{Film_Skyrmion_Crystal} X. Z. Yu, Y. Onose, N. Kanazawa,
J. H. Park, J. H. Han, Y. Matsui, N. Nagaosa, and Y.
Tokura, Nature {\bf 465}, 901 (2010).
%
%
\bibitem{Current_Driven_Motion1} J. C. Slonczewski, J. Magn. Magn. Mater. {\bf 159}, L1 (1996).
\bibitem{DomainWall_Dynamics_Review} G. Tatara, H. Kohno,
and J. Shibata, Phys. Rep. {\bf 468}, 213 (2008).
\bibitem{Collective_Coordinate} O. A. Tretiakov, D. Clarke, G.-W. Chern, Ya. B. Bazaliy, and O.
Tchernyshyov, Phys. Rev. Lett. {\bf 100}, 127204 (2008).

\bibitem{Berry_Phase_Electricity} S. E. Barnes and S. Maekawa, Phys. Rev. Lett.
{\bf 98}, 246601 (2007); S. A. Yang, G. S. D. Beach, C. Knutson,
D. Xiao, Q. Niu, M. Tsoi, and J. L. Erskine, Phys. Rev. Lett. {\bf
102}, 067201 (2009).

\bibitem{Berry_Phase_Electricity_Feedback}
J. Foros, A. Brataas, Y. Tserkovnyak, and G. E. W. Bauer, Phys.
Rev. B {\bf 78}, 140402 (R) (2008); Y. Tserkovnyak and C. H. Wong,
Phys. Rev. B {\bf 79}, 014402 (2009); S. Zhang and S. S.-L. Zhang,
Phys. Rev. Lett. {\bf 102}, 086601 (2009).

\bibitem{Wong:09}
C. H. Wong and Y. Tserkovnyak, Phys. Rev. B {\bf 80}, 184411
(2009).

\bibitem{Wong:10}
C. H. Wong and Y. Tserkovnyak, Phys. Rev. B {\bf 81}, 060404(R)
(2010).

\bibitem{Qi_TI}
X.-L. Qi, T. L. Hughes, and S.-C. Zhang, Phys. Rev. B \textbf{78},
195424 (2008).

\bibitem{Franz_TI} I. Garate and M. Franz, Phys. Rev. Lett. {\bf
104}, 146802 (2010).

\bibitem{Lee_Wen_Nagaosa} P. A. Lee, N. Nagaosa, and X.-G. Wen,
Rev. Mod. Phys. {\bf 78}, 17 (2006).
\bibitem{Kim_Kim} K.-S. Kim and M. D. Kim, Phys. Rev. B
{\bf 75}, 035117 (2007); K.-S. Kim and M. D. Kim, Phys. Rev. B
{\bf 77}, 125103 (2008).
\bibitem{ye}
J. Ye, Y. B. Kim, A. J. Millis, B. I. Shraiman, P. Majumdar, and
Z. Te\u{s}anovi\'{c}, Phys. Rev. Lett. \textbf{83}, 3737 (1999).
\bibitem{AHE} N. Nagaosa, J. Sinova, S. Onoda, A. H. MacDonald, and N. P.
Ong, Rev. Mod. Phys. {\bf 82}, 1539 (2010).



\bibitem{NLSM_Hopf} A. G. Abanov and P. B. Wiegmann, Nucl. Phys. B {\bf
570} 685 (2000).
\bibitem{Soliton_TextBook} R. Rajaraman, \textit{Solitons and
Instantons}(Elsevier Science, New York, 2003).


\bibitem{Index_Theorem} C. Nash, \textit{Differential Topology and
Quantum Field Theory} (Elsevier Science, San Diego, 2004).
\bibitem{Sinova} J. Sinova, D. Culcer, Q. Niu, N. A. Sinitsyn, T. Jungwirth, and A. H.
MacDonald, Phys. Rev. Lett. {\bf 92}, 126603 (2004).

\bibitem{Shibata} J. Shibata, Y. Nakatani, G. Tatara, H. Kohno,
and Y. Otani, Phys. Rev. B {\bf 73}, 020403(R) (2006).


\bibitem{Heun}
S. Yu. Slavyanov, in {\it Heun's Differential Equations}, ed. By
A. Ronveaux, Part B (Oxford, Clarendon, 1995).
\end{thebibliography}
\end{document}